\documentclass{siamart220329}
\usepackage{makeidx}
\usepackage{multirow}
\usepackage{multicol}
\usepackage{graphicx}
\usepackage{epstopdf}
\usepackage{ulem}
\usepackage{hyperref}
\usepackage{amsmath}
\usepackage{amssymb}
\usepackage{setspace}
\usepackage{color}
\usepackage{float}
\author{Aishwarjya Gogoi}
\title{}
\usepackage[paperwidth=612pt,paperheight=792pt,top=72pt,right=72pt,bottom=72pt,left=72pt]{geometry}

\makeatletter
	{\par\setlength{\parindent}{#3}
	\setlength{\leftmargin}{#1}       \setlength{\rightmargin}{#2}%
	\advance\linewidth -\leftmargin       \advance\linewidth -\rightmargin%
	\advance\@totalleftmargin\leftmargin  \@setpar{{\@@par}}%
	\parshape 1\@totalleftmargin \linewidth\ignorespaces}{\par}%
\makeatother 

\begin{document}


\title{A low diffusion HLL-CPS scheme for all Mach number flows\thanks{Submitted to SIAM Journal on Scientific Computing}}

\author{A. Gogoi and J. C. Mandal \thanks{Department of Aerospace Engineering, Indian Institute of Technology, Bombay,
Mumbai, 400076, India}}

\maketitle
\begin{abstract}
A low diffusion version of the HLL-CPS scheme for resolving the shear layers and the flow features at low Mach numbers is presented here. The low diffusion HLL-CPS scheme is obtained by reconstructing the velocities at the cell interface with the face normal Mach number and a pressure function. Asymptotic analysis of the modified scheme shows a correct scaling of the pressure at low Mach numbers and a significant reduction in numerical dissipation. The robustness of the HLL-CPS scheme for strong shock is improved by reducing the contribution of the contact wave in the vicinity of the shock. The improvement in robustness for strong shock is demonstrated analytically through linear perturbation and matrix stability analyses. A set of numerical test cases are solved to demonstrate the efficacy of the proposed scheme over a wide range of Mach numbers. 
\end{abstract}
\section{Introduction }
The approximate Riemann solvers are quite popular for computing the convective fluxes in the Euler and Navier-Stokes equations because of their ability to resolve the non-linear and linearly degenerate discontinuities. Examples of such approximate Riemann solvers are the Roe scheme \cite{roe}, the Osher scheme \cite{osher}, and the HLL-family of schemes like HLL \cite{hll}, HLLE \cite{einf1}, HLLEM \cite{einf2}, HLLC \cite{toro2, batt}. The Roe, Osher, HLLEM, and HLLC schemes incorporate all the linearly degenerate and non-linear waves that are present in the Riemann problem of the Euler equations. Thus, they are categorized as complete approximate Riemann solvers. On the other hand, the HLL \cite{hll}, HLLE \cite{einf1}, and HLL-CPS \cite{mandal} schemes are categorized as incomplete approximate Riemann solvers as these schemes omit one or more waves present in the Riemann problem. Among these approximate Riemann solvers, the HLL-family schemes are capable of resolving isolated shocks exactly and possess positivity-preserving and entropy-satisfying properties desirable for a numerical scheme.As a result,the complete or all-wave solvers of the HLL family, like the HLLEM \cite{einf2} and HLLC \cite{toro2, batt} schemes, are widely used in the simulation of compressible flows. However, the approximate Riemann solvers encounter difficulties in resolving the flow features at very low Mach numbers. Also, the complete Riemann solvers are prone to numerical shock instabilities like carbuncle phenomenon, odd-even decoupling, kinked Mach stem, low-frequency post-shock oscillations, etc. at very high speeds.

The approximate Riemann solvers are known to produce non-physical discrete results in low Mach number flows. Several different approaches like preconditioning \cite{guillard1, park-pre}, eigenvalue modification \cite{li-roe}, velocity reconstruction \cite{thorn1}, velocity jump scaling \cite{rieper1,rieper2, ossw}, momentum jump scaling \cite{li2}, central differencing of momentum flux \cite{della1}, the addition of anti-diffusion terms \cite{della2}, etc., have been developed to alleviate the non-physical solutions and enhance the performance of these schemes at low Mach numbers. It has been shown by Guillard and Viozat \cite{guillard1} and Rieper \cite{rieper1, rieper2} that, in the limit as Mach number tends to zero, the pressure fluctuations are of the order of Mach number in the Roe scheme for the discrete Euler equations, whereas it is of the order of Mach number squared for the continuous Euler equations. Guillard and Viozat have \cite{guillard1} proposed a preconditioned Roe scheme that has a correct pressure scaling with the Mach number. Park and Kwon \cite{park-pre} have proposed a preconditioned HLLE+ scheme for low Mach number flows. Li and Gu \cite {li-roe} have observed that the preconditioned Roe schemes suffer from the global cut-off Mach number problem. They \cite{li-roe} have also proposed an all-speed Roe scheme by modifying the non-linear eigenvalues in the numerical dissipation terms \cite{li-roe}. Thornber et al. \cite{thorn1} have proposed a modification of the reconstructed velocities at the cell interface of the Godunov-type schemes like HLLC \cite{toro2} such that the velocity difference reduces with decreasing Mach number and the arithmetic mean is reached at zero Mach number. Thornber et al. \cite{thorn1} have shown both analytically and numerically that the numerical dissipation of their scheme becomes constant in the limit of zero Mach number, while the numerical dissipation of the traditional schemes tends to infinity in the limit of zero Mach number. 
Rieper \cite{rieper1, rieper2} has proposed a low Mach number fix for the Roe scheme comprising of scaling of the normal velocity jump in the numerical dissipation terms by a local Mach number function. With the low Mach number fix for the Roe scheme, Rieper has demonstrated accurate results for inviscid flow over a cylinder at very low Mach numbers. The low Mach number fix of Rieper has been applied to both the normal and tangential velocity jumps in the numerical dissipation terms of the Roe scheme by Osswald et al. \cite{ossw}. Osswald et al. have demonstrated a reduction in numerical dissipation in low Mach number flows with accuracy comparable to Thornber et al.'s method. Dellacherie \cite{della1} has observed that the Godunov-type schemes cannot be accurate at low Mach numbers due to loss of invariance. Dellacherie has proposed central differencing to discretize the momentum flux or to discretize the pressure gradient in momentum flux to recover invariance property and improve accuracy in low Mach number flows. Dellacherie et al. \cite{della2} have proposed stable, all Mach Godunov-type schemes by adding Mach number-based anti-diffusion terms to the momentum flux in the Godunov-type schemes such that central-differencing is obtained in the limit of Mach zero.

It may be noted that for high-speed flows, the approximate Riemann solvers that resolve the linearly degenerate intermediate waves, like the Roe \cite{roe}, HLLEM \cite{einf1}, and HLLC \cite{toro2} schemes are susceptible to numerical shock instabilities like odd-even decoupling, kinked Mach stem, and carbuncle phenomena. On the other hand, the HLLE scheme \cite{hll, einf1} which does not resolve the linearly degenerate intermediate waves, is free from numerical shock instability problems. Schemes like HLL-CPS \cite{mandal} which resolve one of the intermediate waves, namely the contact wave, have been shown to be free from numerical shock instabilities. However, the HLL-CPS scheme is susceptible to numerical shock instabilities under certain conditions \cite{sangeet-hllcps}. 
The cause of numerical instabilities in the complete Riemann solvers has been investigated by several authors. Quirk \cite{quirk} has carried out linearized perturbation analysis and observed that the schemes in which the pressure perturbations feed the density perturbations are afflicted by the odd-even decoupling problem. Quirk \cite{quirk} has proposed a hybrid Roe-HLLE scheme to overcome the numerical instability problems of the Roe scheme. Sanders et al. \cite{sand} have observed that the instability is a result of inadequate cross-flow dissipation offered by the strictly upwind schemes and proposed a multi-dimensional upwind dissipation to eliminate instability in the Roe scheme. Liou \cite{liou1} has conjectured that the schemes that suffer from shock instability contain a pressure jump term in the numerical mass flux while those free from shock instability do not contain a pressure jump term in the mass flux. 
Pandolfi and D'Ambrosio \cite{pand} have carried out improved perturbation analysis of various numerical schemes by including the velocity perturbations. They \cite{pand} have concluded that the schemes in which the pressure perturbations create density perturbations are strong carbuncle-prone schemes. Dumbser et al. \cite{dumb-matrix} have developed a matrix stability method to evaluate the stability of the numerical schemes and have observed that the origin of the carbuncle phenomenon is localized in the upstream region of the shock and the location of the intermediate point in the numerical shock structure play a key role in the onset of the instability. Several rotated Riemann solvers \cite{ren, nishi, huang} and hybrid HLLC-HLL \cite{huang, hllc-hll} schemes have been proposed to overcome the numerical shock instability problems of the contact-capturing and shear wave-resolving schemes. Shen et al. \cite{shen} have identified the exact resolution of the shear wave as the cause of the numerical shock instabilities in the HLLC scheme and have proposed a hybrid, contact-capturing but shear dissipative HLLC-HLLCM scheme to overcome the instability problems in the HLLC scheme. Similarly, Xie et al. \cite{xie-hllem} have identified the shear wave as the cause of instability in the HLLEM Riemann solver and have introduced a pressure switch in the HLLEM Riemann solver such that the anti-diffusion terms corresponding to the shear wave are reduced in the vicinity of the shock. Kemm \cite{kemm} has shown that the resolution of the entropy wave in the HLLEM scheme also contributed to the numerical shock instabilities, but the contribution of the entropy wave is small as compared to the contribution of the shear wave. Simon and Mandal \cite{simon1, simon2, simon-hllem} have also proposed strategies like anti-diffusion control (ADC) and Selective Wave-speed Modification (SWM) to cure the numerical instability problems in the HLLC and HLLEM Riemann solvers. In recent years, alternate formulations within the HLLE framework have also been proposed to resolve the linearly degenerate discontinuities while avoiding the numerical shock instability problems. In the HLL-CPS scheme proposed by Mandal and Panwar \cite{mandal}, the flux is split into convective and pressure parts and evaluated separately. The convective flux is evaluated with a simple upwind method while the pressure flux is evaluated with the HLL method where the density jumps are replaced by corresponding pressure jumps based on isentropic flow assumption for the resolution of the contact wave. The HLL-CPS scheme \cite{mandal} is free from numerical shock instabilities like odd-even decoupling, carbuncle phenomenon, kinked Mach stem, etc. The HLL-CPS scheme has been shown by Gogoi et al. \cite{gogoi} to be more stable than similar contact-capturing schemes like HLLCM \cite{shen} and HLLEC \cite{xie-hllem}. However, the scheme exhibits mild shock instability for the supersonic corner test case. Moreover, the HLL-CPS scheme is not efficient in resolving shear layers \cite{xie-hllem, kita1, kita-slau, hu, tan}. Kitamura \cite{kita1,kita-slau} has shown that although the HLL-CPS scheme preserves contact discontinuity exactly in 1D, the scheme is incapable of accurately reproducing Blasius' analytical velocity profile for a Mach 0.20 laminar flow over a flat plate. Hu et al. \cite{hu} have shown that the numerical dissipation of the HLL-CPS scheme across an inviscid shear wave is not zero and has attributed the inefficiency of the HLL-CPS scheme to its non-zero numerical dissipation. Tan et al. \cite{tan} have observed that the schemes like HLL-CPS fail to obtain accurate solutions for viscous flow problems and other shear-dominated flow problems due to their high dissipative behaviour for contact discontinuities.

In the field of aeronautics, there are many situations where both incompressible flow regimes with very low Mach numbers and compressible flow regimes with shocks coexist at the same time. Thus, many all Mach number AUSM-family \cite{shima, chen1}, HLLC-type \cite{xie-hllc, chen-hllc}, Roe-type \cite{li-roe, xie-roe} and HLLEM-type \cite{qu-hllems} schemes have been proposed in recent years to handle the mixed high and low speed flow problems. In these all Mach number schemes, fixes for resolving low Mach numbers flow features and suppression of numerical shock instabilities are implemented. For example, in the all Mach number HLLC scheme proposed by Xie et al. \cite{xie-hllc} and the Roe scheme proposed by Xie et al. \cite{xie-roe}, the fix of Dellacherie et al. for Godunov-type schemes \cite{della2} is used to improve low Mach number accuracy while the contribution of the pressure jump terms is reduced to suppress the numerical shock instabilities at high speeds. In the all Mach number HLLEMS-AS scheme proposed by Qu et al. \cite{qu-hllems}, accuracy at low Mach numbers is achieved by scaling the velocity jumps in the momentum equation with a Mach number function. 

In the present work, the HLL-CPS scheme is enhanced to an all Mach number scheme by improving its robustness for high-speed flows and by improving its ability to resolve shear layers and flow features at low Mach numbers. A velocity reconstruction method based on the face normal Mach number and a pressure function is proposed in the HLL-CPS scheme for efficient resolution of shear layers and low Mach flow features. The paper is organized into seven sections. In Sections 2 and 3, the governing equations and their finite volume discretization are described. In Section 4, the HLL-CPS scheme for the inviscid flux is presented and the need for low Mach corrections is established through asymptotic analysis and numerical Reynolds number evaluation. In Section 5, an improved HLL-CPS scheme is proposed along with appropriate analyses to demonstrate the suitability of the proposed scheme for all Mach number flows. In Section 6, numerical results for test cases are presented for speeds ranging from very low Mach number to very high Mach number. Finally, conclusions are presented in Section 7.

\section{Governing equations}
The governing equations for the two-dimensional inviscid  compressible flow can be written in their conservative form as
\begin{equation} \label{2dge}
\dfrac{\partial{}\mathbf{\acute{U}}}{\partial{}t}+\dfrac{\partial{}\mathbf{\acute{F}(\acute{U})}}{\partial{}x}+\dfrac{\partial{}\mathbf{\acute{G}(\acute{U})}}{\partial{}y}=0
\end{equation}
where $\mathbf{\acute{U}}$, $\mathbf{\acute{F}(\acute{U})}$ and $\mathbf{\acute{G}(\acute{U})}$ are the vector of conserved variables, x-directional and y-directional fluxes respectively  defined as
\begin{equation}
\mathbf{\acute{U}}=\left[\begin{array}{c} \rho \\ \rho{}u \\ \rho{}v \\ \rho{}E \end{array}\right] \hspace{1cm} \mathbf{\acute{F}(\acute{U})}=\left[\begin{array}{c} \rho{}u \\ \rho{}u^2+p \\ \rho{}uv \\ (\rho{}E+p)u \end{array}\right]  \hspace{1cm} \mathbf{\acute{G}(\acute{U})}=\left[\begin{array}{c} \rho{}v \\ \rho{}uv \\ \rho{}v^2+p \\ (\rho{}E+p)v \end{array}\right] 
\end{equation}
where $\rho$, $u$, $v$, $p$, and $E$ stand for density, x-directional and y-directional velocities in global coordinates, pressure, and specific total energy. The system of equations can be closed by the equation of state
\begin{equation}
p=(\gamma-1)\left(\rho{}E-\dfrac{1}{2}\rho(u^2+v^2)\right)
\end{equation} 
where $\gamma$ is the ratio of specific heat. In this paper, a calorically perfect gas with $\gamma=1.4$ is considered.
The integral form  of the governing equations is
\begin{equation}\label{2dns}
	\frac{d}{dt}\int_{\Omega{}}\mathbf{\acute{U}}d\Omega{}+\oint_{\partial{}\Omega{}}(\mathbf{\acute{F},\acute{G}}).\mathbf{n}ds=0
\end{equation}
where $\partial{}\Omega$ is the surface enclosing the control volume  $\Omega$ and  $\mathbf{n}$ denotes the outward pointing unit normal vector to the surface $\partial\Omega$.

\section{Finite volume discretization}
The finite volume discretization of equation (4) for a structured, quadrilateral mesh is
\begin{equation}
\dfrac{d\mathbf{\acute{U}}_i}{dt}=-\dfrac{1}{|\Omega|_i}\sum_{l=1}^4[(\mathbf{\acute{F}},\mathbf{\acute{G}})_l.\mathbf{n}_l]\Delta{}s_l
\end{equation}
where $\mathbf{\acute{U}}_i$ is the cell averaged conserved state vector, $\mathbf{n}_l$ denotes the unit surface normal vector and $\Delta{}s_l$ denotes the length of each interface. In order to apply the Riemann solver for flux calculation, the rotational invariance property of the Euler equations is utilized to express the flux $(\mathbf{\acute{F},\acute{G})}_l.\mathbf{n}_l$ as 
\begin{equation} \label{euler-equation}
\dfrac{d\mathbf{\acute{U}}_i}{dt}=-\dfrac{1}{|\Omega|_i}\sum_{l=1}^{4}(\mathcal{T}_{il})^{-1}\mathbf{F}(\mathcal{T}_{il}\mathbf{\acute{U}}_i,\mathcal{T}_{il}\mathbf{\acute{U}}_l)\Delta{}s_l
\end{equation}
where  $\Delta{}s_l$ is the edge length,  $\mathbf{F}(\mathcal{T}_{il}\mathbf{\acute{U}}_i,\mathcal{T}_{il}\mathbf{\acute{U}}_l)$  is the inviscid  face normal flux vector, $\mathcal{T}_{il}$ is the rotation matrix and ${\mathcal{T}_{il}}^{-1}$ is its inverse at the edge between cell $i$ and its neighbour $l$. The rotation matrix and its inverse are respectively
 \begin{equation}
\mathcal{T}_{il}=\left[\begin{array}{ccccc}1& 0& 0& 0\\ 0& n_x& n_y& 0 \\  0& -n_y& n_x& 0 \\ 0& 0& 0& 1\end{array}\right] \hspace{1cm} \text{and} \hspace{1cm} {\mathcal{T}_{il}}^{-1}=\left[\begin{array}{ccccc}1& 0& 0& 0\\ 0& n_x&- n_y& 0 \\  0& n_y& n_x& 0 \\ 0& 0& 0& 1\end{array}\right]
\end{equation}
where $\mathbf{n}=(n_x,n_y)$ is the outward unit normal vector on the edge between cells $i$ and its neighbour $l$.

In the next section, the HLL-CPS scheme for evaluating the inviscid face normal flux  $\mathbf{F}(\mathcal{T}_{il}\mathbf{\acute{U}}_i,\mathcal{T}_{il}\mathbf{\acute{U}}_l)$ shown in equation (\ref{euler-equation}) is described.

\section{Analysis of the original HLL-CPS scheme}
In the HLL-CPS scheme \cite{mandal} the inviscid flux is split into convective and pressure parts following either Zha-Bilgen formulation \cite{zha1} or AUSM formulation \cite{ausm1}, before discretization. A simple upwind scheme is used for evaluating the convective part of the flux and the HLL method \cite{hll} with the density jumps replaced by the corresponding pressure jumps is used for evaluating the pressure flux.
\subsection{Inviscid flux of the original HLL-CPS scheme}
The  inviscid HLL-CPS flux \cite{mandal} at the interface of cell $i$ and its neighbour $l$ is
\begin{equation} \label{hllcps-total}
\begin{split}
& \mathbf{F}(\mathcal{T}_{il}\mathbf{\acute{U}}_i,\mathcal{T}_{il}\mathbf{\acute{U}}_l)=  M_{nK}\left[\begin{array}{c}\rho{} \\ \rho{}u_n \\ \rho{}u_t \\ \rho{}E\end{array}\right]_Ka_K+ 
\dfrac{1}{2}\left[\begin{array}{c} 0 \\ p_L+p_R \\ 0 \\ (pu_n)_L+(pu_n)_R\end{array}\right]   + \\ &\dfrac{S_R+S_L}{2(S_R-S_L)}\left[\begin{array}{c} 0 \\ p_L-p_R \\ 0 \\ (pu_n)_L-(pu_n)_R\end{array}\right] -  \dfrac{S_RS_L}{\bar{a}^{2}(S_R-S_L)}\left[\begin{array}{c}p_L-p_R\\(pu_n)_L-(pu_n)_R\\(pu_t)_L-(pu_t)_R\\ \dfrac{\bar{a}^{2}}{\gamma{}-1}(p_L-p_R)+\dfrac{1}{2}((pq^{2})_L-(pq^{2})_R)\end{array}\right]
\end{split}
\end{equation}
where $u_n=un_x+vn_y$ and $u_t=-un_y+vn_x$ are the normal and tangential velocities respectively, $q^2=u_n^2+u_t^2$ is the absolute velocity, $S_R$ and $S_L$ are the fastest right and left running wave speeds. Here
\begin{equation} \label{mnk}
M_{nK}=\left\lbrace\begin{array}{c}\dfrac{\bar{u}_n}{\bar{u}_n-S_L} \hspace{4mm} \text{if} \hspace{2mm} \bar{u}_n\geq0.0\\
\dfrac{\bar{u}_n}{\bar{u}_n-S_R} \hspace{4mm} \text{if} \hspace{2mm} \bar{u}_n<0.0\end{array}\right.
\end{equation}
\begin{equation} \label{ak}
a_l=\left\lbrace\begin{array}{c} \ u_{nL}-S_L \hspace{4mm} \text{if} \hspace{2mm} \bar{u}_n\geq0.0\\\ u_{nR}-S_R \hspace{4mm} \text{if} \hspace{2mm} \bar{u}_n < 0.0\end{array}\right.
\end{equation}
\begin{equation}\label{flux1-upwind-k}
K=\left\lbrace\begin{array}{c} L \hspace{4mm} \text{if} \hspace{4mm} \bar{u}_n\geq0.0\\ R \hspace{4mm} \text{if} \hspace{4mm} \bar{u}_n < 0.0\end{array}\right.
\end{equation}
where $\bar{u}_n=\frac{(u_{nL}+u_{nR})}{2}$ is the average normal velocity at the cell interface. The wave speeds are defined as
\begin{equation} \label{wavespeed}
\begin{array}{l}
S_L=min(0,\;u_{nL}-a_L,\;\tilde{u}_n-\tilde{a}) \hspace{1cm} S_R=max(0,\;u_{nR}+a_R,\;\tilde{u}_n+\tilde{a})
\end{array}
\end{equation}
where $\tilde{u}_n$ and $\tilde{a}$ are the Roe-averaged normal velocity and speed of sound respectively at the interface. 

The ability of the HLL-CPS scheme to resolve low Mach number flow features and shear layers is assessed next using techniques like asymptotic analysis and numerical Reynolds number evaluation.
\subsection{Asymptotic analysis of the original HLL-CPS scheme}
In the present work, asymptotic analysis of the HLL-CPS scheme is carried out like Guillard and Viozat \cite{guillard1} to determine whether the pressure fluctuation in the scheme is consistent with the continuous Euler equation. In the asymptotic analysis, a Cartesian grid of uniform mesh size is considered where $\mathbf{i}=(i,j)$ is the index of the cell and $\upsilon(\mathbf{i})=((i+1,j)$, $(i-1,j)$, $(i,j+1)$, $(i,j-1))$ are the neighbours of the cell $\mathbf{i}$. For low-speed flows, the wave speeds can be approximated as $S_R=u_n+a$ and $S_L=u_n-a$. Therefore, the following simplification in the terms appearing in equation (\ref{hllcps-total}) can be made:
\begin{equation}
\dfrac{S_RS_L}{(S_R-S_L)}=\dfrac{u_n^2-a^2}{2a}, \hspace{1cm} \dfrac{S_R+S_L}{S_R-S_L}=\dfrac{u_n}{a}, \hspace{1cm} M_{nK}a_l=\dfrac{u_n(u_{nK}-S_K)}{u_{n}-S_K}=u_n
\end{equation}
At low Mach numbers, the HLL-CPS scheme shown in equations (\ref{euler-equation}) and (\ref{hllcps-total})  can thus be written as 
\begin{equation}\label{hllcps-lowspeed}
\begin{split}
& |\Omega|_i\dfrac{\partial{}\bar{U}_i}{\partial{}t}+ \sum_{l\epsilon\upsilon(i)}\mathcal{T}_{il}^{-1}\left[u_{n,il}\left[\begin{array}{c} \rho\\ \rho{}u_n\\ \rho{}u_t \\ \rho{}E\end{array}\right]_K+\left[\begin{array}{c} 0 \\ p\\ 0 \\ pu_n\end{array}\right]_{il}+ \left(\dfrac{u_n}{2a}\right)_{il}\left[\begin{array}{c} 0 \\\Delta_{il}(p)\\ 0 \\ \Delta_{il}(pu_n)\end{array}\right]\right]\Delta{}s_{il}-\\ &\mathcal{T}_{il}^{-1}\left(\dfrac{u_n^2-a^2}{2a^3}\right)_{il}\left[\begin{array}{c}\Delta_{il}(p) \\ \Delta_{il}(pu_n) \\ \Delta_{il}(pu_t) \\ \dfrac{a^2_{il}\Delta_{il}(p)}{\gamma-1}+\dfrac{\Delta_{il}(pq^2)}{2} \\\end{array}\right]\Delta{}s_{il}=0
\end{split}
\end{equation}
where $\Delta_{il}=(.)_L-(.)_R$. The flow variables are non-dimensionalized as 
\begin{equation}
\bar{\rho}=\dfrac{\rho}{\rho{}_*} \hspace{1cm} \bar{u}=\dfrac{u}{u_*} \hspace{1cm} \bar{p}=\dfrac{p}{\rho_*a_*^2} \hspace{1cm} \bar{E}=\dfrac{E}{a_*^2}\hspace{1cm}  \bar{t}=\dfrac{tu_*}{s_*} \hspace{1cm} \bar{\Omega}=\dfrac{\Omega}{s_*^2}
\end{equation}
where $\rho_*$ is the reference density, $u_*$ is the reference velocity, $a_*$ is the reference speed of sound, and $s_*$ is the reference length. 

The HLL-CPS scheme shown in (\ref{hllcps-lowspeed}) can be re-written in the non-dimensional form in the Cartesian frame as 
\begin{equation} \label{hllcps-lowspeed-after-rotation}
\begin{split}
& |\bar{\Omega}|\dfrac{\partial{}\bar{U}_i}{\partial{}\bar{t}}+  \sum_{l\epsilon\upsilon(i)}\bar{u}_{n,il}\left[\begin{array}{c} \bar{\rho}\\ \bar{\rho}\bar{u}\\ \bar{\rho}\bar{v} \\ \bar{\rho}\bar{E}\end{array}\right]_K+\left[\begin{array}{c} 0 \\ \dfrac{\bar{p}}{M_*^2}n_x\\  \dfrac{\bar{p}}{M_*^2}n_y \\ \bar{p}\bar{u_n}\end{array}\right]_{il}+ \left(\dfrac{\bar{u}_nM_*}{2\bar{a}}\right)_{il}\left[\begin{array}{c} 0 \\ \left(\dfrac{n_x}{M_*^2}\right)_{il}\Delta_{il}(\bar{p})\\  \left(\dfrac{n_y}{M_*^2}\right)_{il} \Delta_{il}(\bar{p})\\ \Delta_{il}(\bar{p}\bar{u}_n)\end{array}\right]\Delta{}s_{il}-  \\ &  \sum_{l\epsilon\upsilon(i)}\left(\dfrac{\bar{u}_n^2M_*}{2\bar{a}^3}-\dfrac{1}{2\bar{a}M_*}\right)_{il}\left[\begin{array}{c}\Delta_{il}(\bar{p}) \\ \Delta_{il}(\bar{p}\bar{u}) \\ \Delta_{il}(\bar{p}\bar{v}) \\ \dfrac{\bar{a}^2_{il}\Delta_{il}(\bar{p})}{\gamma-1}+\dfrac{\Delta_{il}(\bar{p}\bar{q}^2)M_*^2}{2} \\\end{array}\right]\Delta{}s_{il}=0
\end{split}
\end{equation}
The flow variables are expanded using the following asymptotic expansions 
\begin{equation} \label{expansion}
\bar{\rho}=\bar{\rho_0}+M_*\bar{\rho_1}+M_*^2\bar{\rho_2}, \hspace{1cm} \bar{u}_n=\bar{u}_{n0}+M_*\bar{u}_{n1}+M_*^2\bar{u}_{n2}, \hspace{1cm} \bar{p}=\bar{p}_0+M_*\bar{p}_1+M_*^2\bar{p}_2 
\end{equation}
Expanding and re-arranging terms in the continuity and momentum equations in the equal power of $M_*$, we obtain the following:
\begin{itemize}
\item {Order of $M_*^0$}
\begin{itemize}
\item {continuity equation}
\begin{equation}
|\bar{\Omega}|\dfrac{\partial\bar{\rho}_{0i}}{\partial{}\bar{t}}+\sum_{l\epsilon\upsilon(i)}\left(\bar{u}_{n0,il}\bar{\rho}_{0K}+\dfrac{1}{2\bar{a}_{0,il}}\Delta_{il}\bar{p}_1\right)\Delta{}s_{il}=0
\end{equation}
Here, the first term in the flux is the physical flux while the second term is the numerical dissipation. The pressure jump term $\Delta_{il}p_1$ contributes to the numerical dissipation of the scheme.
\item{x-momentum equation}
\begin{equation} \label{x-mom-om0-hllcps}
\begin{split}
& |\bar{\Omega}|\dfrac{\partial(\bar{\rho}_0\bar{u}_0)_i}{\partial{}\bar{t}}+  \sum_{l\epsilon\upsilon(i)}\left(\bar{u}_{{n0il}}(\bar{\rho}_{0}\bar{u}_0)_{K}+(\bar{p}_2n_x)_{il}\right)\Delta{}s_{il} \\ & +\sum_{l\epsilon\upsilon(i)}\left(\left(\dfrac{\bar{u}_{n0}}{2\bar{a}_0}n_x\right)_{il}\Delta_{il}\bar{p}_1+\dfrac{1}{2\bar{a}_{0,il}}\Delta_{il}(\bar{p_1}\bar{u_0})+\Delta_{il}(\bar{p_0}\bar{u_1})\right)\Delta{}s_{il}=0
\end{split}
\end{equation}
Here, the first two terms in the flux part of the equation are the physical fluxes while the remaining terms are the numerical dissipation terms. The pressure and velocity jump terms $\Delta_{il}p_0$, $\Delta_{il}p_1$, $\Delta_{il}u_0$ and $\Delta_{il}u_1$ contribute to the numerical dissipation of the scheme.
\item{y-momentum equation} 
\begin{equation} \label{y-mom-om0-hllcps}
\begin{split}
& |\bar{\Omega}|\dfrac{\partial(\bar{\rho}_0\bar{v}_0)_i}{\partial{}\bar{t}}+  \sum_{l\epsilon\upsilon(i)}\left(\bar{u}_{{n0,il}}(\bar{\rho}_{0}\bar{v}_{0})_K+(\bar{p}_2n_y)_{il}\right)\Delta{}s_{il}\\ &
+\sum_{l\epsilon\upsilon(i)}\left(\left(\dfrac{\bar{u}_{n0}}{2\bar{a}_0}n_y\right)_{il}\Delta_{il}\bar{p}_1+\dfrac{1}{2\bar{a}_{0,il}}(\Delta_{il}(\bar{p_1}\bar{v_0})+\Delta_{il}(\bar{p_0}\bar{v_1})\right)\Delta{}s_{il}=0
\end{split}
\end{equation}
\end{itemize}
\item {Order of $M_*^{-1}$} 
\begin{itemize}
\item {continuity equation}
\begin{equation}\label{deltap0}
\sum_{l\epsilon\upsilon(i)}\dfrac{1}{2\bar{a}_{0,il}}\Delta_{il}\bar{p}_0=0
\end{equation}
\item{x-momentum equation}
\begin{equation} \label{x-mom-om-minus1}
\sum_{l\epsilon\upsilon(i)}\left((\bar{p}_1n_x)_{il}+\left(\dfrac{\bar{u}_{n0}}{2\bar{a}_0}n_x\right)_{il}\Delta_{il}\bar{p}_0+\dfrac{1}{2\bar{a}_{0,il}}(\bar{p}_{0,il}\Delta_{il}\bar{u}_{0}+\bar{u}_{0,il}\Delta_{il}\bar{p}_0)\right)=0
\end{equation}
\item{y-momentum equation}
\begin{equation} \label{y-mom-om-minus1}
\sum_{l\epsilon\upsilon(i)}\left((\bar{p}_1n_y)_{il}+\left(\dfrac{\bar{u}_{n0}}{2\bar{a}_0}n_x\right)_{il}\Delta_{il}\bar{p}_0+\dfrac{1}{2\bar{a}_{0,il}}(\bar{p}_{0,il}\Delta_{il}\bar{v}_{0}+\bar{v}_{0,il}\Delta_{il}\bar{p}_0)\right)=0
\end{equation}
\end{itemize}
\end{itemize}
From the above equation (\ref{deltap0}), we obtain $\bar{p}_0=$constant for $\textbf{i}$. The equations (\ref{x-mom-om-minus1}) and (\ref{y-mom-om-minus1}) therefore reduce to
\begin{equation} \label{pnx}
\sum_{l\epsilon\upsilon(i)}(\bar{p_1}n_x)_{il}=-\sum_{l\epsilon\upsilon(i)}\dfrac{1}{2}\dfrac{\bar{p}_{0,il}}{\bar{a}_{0,il}}\Delta_{il}\bar{u}_0
\end{equation}
\begin{equation}\label{pny}
\sum_{l\epsilon\upsilon(i)}(\bar{p_1}n_y)_{il}=-\sum_{l\epsilon\upsilon(i)}\dfrac{1}{2}\dfrac{\bar{p}_{0,il}}{\bar{a}_{0,il}}\Delta_{il}\bar{v}_0
\end{equation}
Rotating the results of x and y momentum equations shown in equations (\ref{pnx}, \ref{pny}) into the face normal direction, we obtain
\begin{equation} \label{p1-normal}
\sum_{l\epsilon\upsilon(i)}\bar{p}_1=-\sum_{l\epsilon\upsilon(i)}\dfrac{1}{2}\dfrac{\bar{p}_{0,il}}{\bar{a}_{0,il}}\Delta_{il}\bar{u}_{n0}
\end{equation}
This implies that $\bar{p}_1\neq{}$ constant for $\textbf{i}$. Therefore, the HLL-CPS scheme permits  pressure fluctuation of the type $p(x,t)=P_0(t)+M_*p_1(x,t)$. Thus, the HLL-CPS scheme is not able to resolve the low Mach number flow features as the pressure fluctuations are not consistent with the continuous Euler equations. It can be seen from equation (\ref{p1-normal}) that the normal velocity jump is the cause of the non-physical behaviour of the HLL-CPS scheme at low Mach numbers.
\subsection{Numerical Reynolds number evaluation of the original HLL-CPS scheme}
The concept of numerical Reynolds number has been introduced by Rieper \cite{rieper1, rieper2} and it is defined as the ratio of the magnitude of convection to the magnitude of artificial dissipation. Rieper has proposed that a first-order upwind scheme for the Euler equations be called \textbf{asymptotically consistent} if all characteristic variables of the corresponding modified equation have a numerical Reynolds number that satisfies $Re_{num}=\mathcal{O}_S\left(\frac{1}{\Delta{}x}\right) \text{ as  M} \to 0 $.  If for a characteristic variable, the modified equation satisfies $ Re_{num}=\mathcal{O}_S\left(\frac{M}{\Delta{}x}\right) \text{ as  M }\to 0$, the scheme can be called \textbf{asymptotically inconsistent} with respect to this characteristic wave. In the present work, the numerical Reynolds number of the HLL-CPS scheme for a pure shear wave is investigated to ascertain the need for improving the shear wave resolution capability of the scheme.

In the special case of a pure shear wave, we have $\rho=const$, $u=const$, $a=const$, and $p=const$. Here, $u$ is the normal velocity and $v$ is the tangential velocity.

To start with, we assume $M<0$ and obtain the momentum flux $\rho{}v$ transverse to the Riemann problem from equation (\ref{hllcps-total})
\begin{equation}
\mathbf{F}^{\rho{}v}=M_{nR}(\rho{}v)_Ra_R-\dfrac{S_RS_L}{a^2(S_R-S_L)}(p_Lv_L-p_Rv_R)
\end{equation}
Since $u=const$ and $a=const$, the following simplifications can be made:

$S_R=u+a,\; S_L=u-a, \; S_RS_L=u^2-a^2, \; S_R-S_L=2a$  and $M_{nR}a_R=u\dfrac{u-S_R}{u-S_R}=u$  

Thus, the fluxes at the interfaces ($i+\frac{1}{2},j$) and $(i-\frac{1}{2},j)$ can be written as:
\begin{equation}
\begin{array}{l}
\mathbf{F}^{\rho{}v}_{i+\frac{1}{2}}=u(\rho{}v)_{i+1}-\dfrac{a^2-u^2}{2a^3}p(v_{i+1}-v_i) \\
\mathbf{F}^{\rho{}v}_{i-\frac{1}{2}}=u(\rho{}v)_i-\dfrac{a^2-u^2}{2a^3}p(v_i-v_{i-1})
\end{array}
\end{equation}
The flux difference is given by
\begin{equation} \label{flux-diff1}
\dfrac{\mathbf{F}^{\rho{}v}_{i+\frac{1}{2}}-\mathbf{F}^{\rho{}v}_{i-\frac{1}{2}}}{\Delta{}x}=\rho{}u\dfrac{\Delta_{i+1,i}(v)}{\Delta{}x}-\dfrac{(a^2-u^2)}{2a^3}p\dfrac{\Delta^2_i(v)}{\Delta{}x}
\end{equation}
where $\Delta_{i+1,i}(v)=v_{i+1}-v_i $ and $\Delta^2_i(v)=v_{i+1}-2v_i+v_{i-1}$. Let \textit{v} be a smooth solution. With the following Taylor expansion about the grid node $x_i$
\begin{equation}
v(x_i+\Delta{}x)=v(x_i)+v_x(x_i)\Delta{}x+\dfrac{1}{2}v_{xx}(x_i)\Delta{}x^2+\mathcal{O}(\Delta{}x^2)
\end{equation}
The term on the Right-Hand side of the equation (\ref{flux-diff1}) leads to the flux difference
\begin{equation}
\rho{}u\dfrac{\Delta_{i+1,i}(v)}{\Delta{}x}-\dfrac{(a^2-u^2)}{2a^3}p\dfrac{\Delta^2_i(v)}{\Delta{}x}=\rho{}u\left(v_x+\dfrac{1}{2}v_{xx}\Delta{}x\right)-\dfrac{a^2-u^2}{2a^3}p(v_{xx}\Delta{}x)
\end{equation}
Using the above equation, along with the result for $M>0$, the modified equation for the transport of a shear wave can be written as
\begin{equation}
\dfrac{\partial}{\partial{}t}\rho{}v+u\dfrac{\partial}{\partial{}x}\rho{}v=\dfrac{1}{2}\dfrac{\partial^2}{\partial{}x^2}(\rho{}v)\Delta{}x\left|u+\left(\dfrac{a^2-u^2}{a}\right)\left(\dfrac{p}{\rho{}a^2}\right)\right|+\mathcal{O}(\Delta{}x^2)
\end{equation}
The numerical Reynolds number for the shear wave can now be calculated as
\begin{equation}
Re_{num,\: HLL-CPS}=\dfrac{|u|}{\left|u+\left(\dfrac{a^2-u^2}{a}\right)\left(\dfrac{p}{\rho{}a^2}\right)\right|\Delta{}x}
\end{equation}
 As $\left|u+\left(\dfrac{a^2-u^2}{a}\right)\left(\dfrac{p}{\rho{}a^2}\right)\right| \to \dfrac{a}{\gamma}$ for $M\to 0$. Therefore,
\begin{equation}
Re_{num,\: HLL-CPS} \to \mathcal{O} \left( \frac{M}{\Delta{}x}\right) as \hspace{1mm} M \to 0
\end{equation}
It may be noted that, for the HLLE scheme \cite{hll, einf1}, the numerical Reynolds number for a pure shear wave is $Re_{num,\:HLLE}=\mathcal{O}(\frac{M}{\Delta{}x}) $ as $M\to 0$ \cite{rieper2}. Therefore, the HLL-CPS scheme shows an increase in the numerical Reynolds number for a pure shear wave over the HLLE scheme by a factor of $\gamma$. It has been shown by Rieper \cite{rieper2} that the numerical Reynolds number of the shear wave resolving schemes like the Roe \cite{roe}, HLLEM \cite{einf2}, and HLLC \cite{toro2} schemes are independent of Mach number, i.e., $Re_{num,\: Roe,\: HLLC,\: HLLEM}=\mathcal{O}\left(\frac{1}{\Delta{x}}\right)$. Thus, it is clear from the above analysis that the HLL-CPS scheme will not be able to accurately resolve shear layers like the Roe, HLLC and HLLEM schemes. It is necessary to modify the HLL-CPS scheme to improve its performance at low Mach numbers and in presence of shear waves. The modification proposed in the HLL-CPS scheme is described in the next section.
\section{Inviscid flux of the proposed HLL-CPS scheme}
In this section, modifications to the original HLL-CPS scheme are proposed for improving its ability to resolve the shear layers and low Mach number flow features. Modifications are also proposed for improving its stability in presence of strong shock.
\subsection {Modified flux for low Mach number and shear flows}
 Several modifications have been proposed to the Godunov-type schemes to improve performance in low Mach number flows. The method of Thornber et al. \cite{thorn1} comprises of reconstruction of the velocities at the cell interface such that the velocity difference reduces with decreasing Mach number and the velocities reach the arithmetic mean at zero Mach number. The velocity reconstruction method  proposed by Thornber et al. \cite{thorn1} has been found to produce accurate results at low Mach numbers when applied to the HLLC scheme \cite{thorn2}. The low Mach number fix proposed by Rieper \cite{rieper2} for the Roe scheme comprises scaling of the normal velocity jump in the numerical dissipation terms by a Mach number function. The low Mach fix of Rieper has been applied to both the normal and tangential velocity jumps in the numerical dissipation terms of the Roe scheme by Osswald et al. \cite{ossw}. It was shown by Osswald et al. that the application of the low Mach correction to both the normal and tangential velocity jumps leads to a reduction in numerical dissipation of  the Roe scheme. The differences between  Rieper's method and Thornber's  method are that the corrections are applied only to the normal velocity jump in the numerical dissipation terms in Rieper's method while the corrections are applied to both the normal and tangential velocities in both the numerical dissipation and physical flux in Thornber's method. In the present work, the velocity reconstruction method of Thornber et al. with suitable modifications is applied to the HLL-CPS scheme. The velocity reconstruction method of Thornber et al. is preferred to the low Mach number fix of Rieper and it will be shown subsequently that the numerical dissipation of the HLL-CPS scheme with Thornber et al.'s method is lower as compared to Rieper's method.
\subsubsection{Velocity reconstruction method for the HLL-CPS scheme}
The reconstructed normal and tangential velocities in the method proposed by Thornber et al. \cite{thorn1} are 
\begin{equation} \label{thornberuv}
\begin{array}{l}
u_{nL}^*=\dfrac{u_{nL}+u_{nR}}{2}+z\dfrac{u_{nL}-u_{nR}}{2} \hspace{1cm} u_{tL}^*=\dfrac{u_{tL}+u_{tR}}{2}+z\dfrac{u_{tL}-u_{tR}}{2} \\
u_{nR}^*=\dfrac{u_{nL}+u_{nR}}{2}+z\dfrac{u_{nR}-u_{nL}}{2} \hspace{1cm}	u_{tR}^*=\dfrac{u_{tL}+u_{tR}}{2}+z\dfrac{u_{tR}-u_{tL}}{2}
\end{array}
\end{equation}
where superscript $^*$ denotes the reconstructed velocities. The Mach number function $z$ is defined as  
\begin{equation} \label{function-z}
z=min\left(max\left(\dfrac{q_L}{a_L},\dfrac{q_R}{a_R}\right),\;1\right)
\end{equation}
The wave speeds are also updated based on the reconstructed velocities. The updated wave speeds are
\begin{equation} \label{wavespeed-lm}
S_L^*=min\left(0,\;u_{nL}^*-a_L,\;\tilde{u}_n^*-\tilde{a}\right) \hspace{1cm}	S_R^*=max\left(0,\;u_{nR}^*+a_R,\;\tilde{u}_n^*+\tilde{a}\right)
\end{equation}
where $\tilde{u}_n^*$ is the reconstructed Roe-averaged face normal velocity.
The HLL-CPS convective and pressure fluxes are evaluated using equation (\ref{hllcps-total}) with the velocities estimated using (\ref{thornberuv}) and wave speeds estimated using  (\ref{wavespeed-lm}).

In the next section, we investigate the suitability of the velocity reconstruction method of Thornber et al. for low Mach number flows by evaluating the numerical Reynolds number of the modified HLL-CPS for a pure shear wave. 
\subsection{Numerical Reynolds number of the modified HLL-CPS scheme}
The numerical Reynolds number of the modified HLL-CPS scheme is evaluated like the HLL-CPS scheme. The momentum flux of the modified HLL-CPS scheme in the direction transverse to the Riemann problem appearing in equations (\ref{hllcps-total}, \ref{thornberuv}, \ref{wavespeed-lm}) can be written as:
\begin{equation}
\begin{array}{l}
\mathbf{F}^{\rho{}v}_{i+\frac{1}{2}}=\rho{}uv_{i+1}^*+\dfrac{a^2-u^2}{2a^3}p(v_{i}^*-v_{i+1}^*) \\
\mathbf{F}^{\rho{}v}_{i-\frac{1}{2}}=\rho{}uv_{i}^*+\dfrac{a^2-u^2}{2a^3}p(v_{i-1}^*-v_{i}^*)
\end{array}
\end{equation}
The flux difference is given by
\begin{equation} \label{flux-diff}
\dfrac{\mathbf{F}^{\rho{}v}_{i+\frac{1}{2}}-\mathbf{F}^{\rho{}v}_{i-\frac{1}{2}}}{\Delta{}x}=\rho{}u\dfrac{\Delta_{i+1,i}(v^*)}{\Delta{}x}-\dfrac{(a^2-u^2)}{2a^3}p\dfrac{\Delta^2_i(v^*)}{\Delta{}x}
\end{equation}
Using  equation (\ref{thornberuv}), the transverse momentum flux difference can be written as 
\begin{equation}
\dfrac{\mathbf{F}^{\rho{}v}_{i+\frac{1}{2}}-\mathbf{F}^{\rho{}v}_{i-\frac{1}{2}}}{\Delta{}x}=\rho{}uz\dfrac{\Delta_{i+1,i}(v)}{\Delta{}x}-\dfrac{(a^2-u^2)}{2a^3}pz\dfrac{\Delta^2_i(v)}{\Delta{}x}
\end{equation}
With the Taylor expansion about grid node $x_i$, the term on the Right-Hand side of the equation (\ref{flux-diff}) leads to the flux difference
\begin{equation}
\rho{}uz\dfrac{\Delta_{i+1,i}(v)}{\Delta{}x}-\dfrac{(a^2-u^2)}{2a^3}pz\dfrac{\Delta^2_i(v)}{\Delta{}x}=  \rho{}uz\left(v_x+\dfrac{1}{2}v_{xx}\Delta{}x\right)-\dfrac{a^2-u^2}{2a^3}pz(v_{xx}\Delta{}x)
\end{equation}
Therefore, we obtain
\begin{equation}
\dfrac{\mathbf{F}^{\rho{}v}_{i+\frac{1}{2}}-\mathbf{F}^{\rho{}v}_{i-\frac{1}{2}}}{\Delta{}x}=\rho{}uz\left(v_{x}+\dfrac{1}{2}v_{xx}\Delta{}x\right)-\dfrac{a^2-u^2}{2a}\dfrac{p}{a^2}z(v_{xx}\Delta{}x) 
\end{equation}
Using the above equation  along with the result for $M>0$, the modified equation for the transport of a shear wave can be written as
\begin{equation}
\dfrac{\partial}{\partial{}t}\rho{}v+u\dfrac{\partial}{\partial{}x}\rho{}v=\dfrac{1}{2}\dfrac{\partial^2}{\partial{}x^2}(\rho{}v)\Delta{}xz\left|u+\left(\dfrac{a^2-u^2}{a}\right)\left(\dfrac{p}{\rho{}a^2}\right)\right|+\mathcal{O}(\Delta{}x^2)
\end{equation}
The numerical Reynolds number for the shear wave can now be calculated as
\begin{equation}
Re_{num}=\dfrac{|u|}{z\left|u+\left(\dfrac{a^2-u^2}{a}\right)\left(\dfrac{p}{\rho{}a^2}\right)\right|\Delta{}x}
\end{equation}
Here, $z=\dfrac{max(q_R,q_L)}{a}$ and $q=\sqrt{u^2+v^2}$. Therefore, 
\begin{equation} \label{re-hllcps-new}
Re_{num} =\dfrac{|u|}{max(q_R,q_L)\left|\dfrac{u}{a}+\left(\dfrac{a^2-u^2}{a^2}\right)\left(\dfrac{p}{\rho{}a^2}\right)\right|\Delta{}x}
\end{equation}
As the Mach number tends to zero, 
$\left|\dfrac{u}{a}+\left(\dfrac{a^2-u^2}{a^2}\right)\left(\dfrac{p}{\rho{}a^2}\right)\right| \to \dfrac{1}{\gamma}$. Therefore, we obtain
\begin{equation}
Re_{num} \to \mathcal{O} \left( \dfrac{\gamma}{\Delta{}x}\right) as \hspace{1mm} M \to 0
\end{equation}
Thus, the numerical Reynolds number of the modified HLL-CPS scheme for a pure shear wave becomes independent of the Mach number as $M\to 0$. However, we observe from equation (\ref{re-hllcps-new}) that the numerical Reynolds number reduces significantly if the absolute velocity is large as compared to the normal velocity ($q>>u$). Further, even though the HLL-CPS scheme with Thornber's low Mach correction is asymptotically consistent for a pure shear wave, the scheme is still unable to resolve the shear wave perfectly like the Roe, HLLEM, and HLLC schemes. Therefore, to resolve the shear wave perfectly, we now re-evaluate the numerical Reynolds number of the scheme by using the face normal Mach number instead  of the local Mach number in Thornber's velocity reconstruction function. The low Mach correction based on the face normal Mach number can be written as
\begin{equation} \label{z-face-normal}
z_{n}=min\left(max\left(\dfrac{u_{nL}}{a_L},\dfrac{u_{nR}}{a_R}\right),1\right)
\end{equation}
For the case of a pure shear wave, Thornber's face normal Mach number-based function becomes $z_n=min\left((u/a),1\right)$, and the numerical Reynolds number for the HLL-CPS scheme shown in equation  (\ref{re-hllcps-new}) becomes
\begin{equation} \label{re-hllcps-tn}
Re_{num} =\dfrac{1}{\left|\dfrac{u}{a}+\left(\dfrac{a^2-u^2}{a^2}\right)\left(\dfrac{p}{\rho{}a^2}\right)\right|\Delta{}x}
\end{equation}
The term in the denominator $ \left|\dfrac{u}{a}+\left(\dfrac{a^2-u^2}{a^2}\right)\left(\dfrac{p}{\rho{}a^2}\right)\right|$ varies from $\dfrac{1}{\gamma}$ at $u=0$ to unity at $u=a$. Considering the term to be close to unity, we obtain
\begin{equation} \label{re-hllcps-tn1}
Re_{num} \approx \dfrac{1}{\Delta{}x}
\end{equation}
Thus, the HLL-CPS scheme with Thornber-type velocity reconstruction based on face normal  Mach number shall be able to resolve the shear wave as accurately as the Roe, HLLEM, and HLLC schemes. However, the Roe, HLLEM, and HLLC schemes are prone to numerical shock instabilities like carbuncle phenomenon, odd-even decoupling, and kinked Mach stem at high Mach numbers. It may be noted that, the HLL-CPS scheme with face normal Mach number-based correction might encounter shock instability problems due to its perfect shear wave resolving ability. Hence, the face normal Mach number-based function is modified to prevent the numerical shock instabilities for high-speed flow problems. The shock-stable low Mach correction function for the HLL-CPS scheme based on the face normal Mach number is defined as

\begin{equation} \label{znp}
z=1-(1-z_n)(1-fp) 
\end{equation} 
where $z_n$ is the face normal Mach number function defined in equation (\ref{z-face-normal}) and $fp$ is the pressure switch defined  as
\begin{equation} \label{fp1}
fp=\sqrt{\dfrac{|p_L-p_R|}{max(p_L,p_R)}}
\end{equation}
To detect a shock, the neighbouring cells must be considered. Therefore, the function $fp$ is calculated as 
\begin{equation} \label{fp2}
\begin{array}{c}
fp_{i+\frac{1}{2},j}=max(fp_{i+\frac{1}{2},j},\;fp_{i,j+\frac{1}{2}}, \;fp_{i,j-\frac{1}{2}}, \;fp_{i+1,j+\frac{1}{2}}, \;fp_{i+1,j-\frac{1}{2}})  \\
fp_{i,j+\frac{1}{2}}=max(fp_{i,j+\frac{1}{2}},\;fp_{i+\frac{1}{2},j}, \;fp_{i+\frac{1}{2},j+1}, \;fp_{i-\frac{1}{2},j}, \;fp_{i-\frac{1}{2},j+1})
\end{array}
\end{equation}

For a pure shear wave, the pressure jump is zero and hence the function $z$ becomes equal to $z_n$. Thus, the scheme can resolve the shear wave perfectly like the Roe, HLLC, and HLLEM schemes. In presence of a shock, the term  $1-fp$ becomes close to zero, the function $z$ becomes close to unity, and the scheme recovers to the robust HLL-CPS scheme.

\subsection {Modified flux for high Mach number flows}
The HLL-CPS scheme is capable of handling high-speed flow problems reasonably well \cite{mandal}. The HLL-CPS scheme is free from the carbuncle phenomenon, odd-even decoupling, and kinked Mach stem problems that are typically encountered by the approximate Riemann solvers that incorporate all the waves in the Riemann problem. However, the HLL-CPS scheme exhibit mild instability for the supersonic corner test case where a Mach 5.09 shock diffracts around a 90-degree corner as described later in Section 6.2.5. Therefore, a need is felt to improve the robustness of the HLL-CPS scheme. It can be seen from the HLL-CPS flux function of equation (\ref{hllcps-total}) that the density jump terms which are present in the classical HLL scheme \cite{hll} are replaced by the corresponding pressure jump terms based on the isentropic flow assumption. The mild instability in the HLL-CPS scheme can be attributed to the total replacement of the density jump terms by pressure jump terms. Therefore, to enhance the robustness of the HLL-CPS scheme we introduce diffusion for the contact wave in the vicinity of a shock by employing a pressure function. The flux function of the improved and robust HLL-CPS scheme can be written as 
\begin{equation} \label{hllcpsr}
\mathbf{F}_{HLL-CPS-RTNP}=\mathbf{F}_{HLL-CPS}+\dfrac{S_RS_L}{S_R-S_L}f_p\left(\Delta{}\rho-\dfrac{\Delta{}p}{\bar{a}^2}\right) \left[1,\;\bar{u}_n,\;\bar{u}_t,\; \dfrac{(\bar{u}_n^2+\bar{u}_t^2)}{2} \right]^T
\end{equation} 
where $\mathbf{F}_{HLL-CPS-RTNP}$ stands for the final modified HLL-CPS flux, $\mathbf{F}_{HLL-CPS}$ is the HLL-CPS flux defined in equation (\ref{hllcps-total}) with reconstructed velocities based on equations (\ref{function-z}, \ref{znp}, \ref{fp1}, \ref{fp2}) and reconstructed wave-speeds (\ref{wavespeed-lm}). In equation \ref{hllcpsr}, $fp$ is the pressure function defined in equations (\ref{fp1}, \ref{fp2}) which controls the amount of dissipation, while the last two terms are the wave strength and right eigenvector of the characteristic field represented by the contact wave. 

The proposed scheme can be referred to as the HLL-CPS-RTNP scheme, that is, the $\mathbf{R}$obust HLL-CPS scheme with $\mathbf{T}$hornber-type low Mach correction based on Face $\mathbf{N}$ormal Mach number and $\mathbf{Pr}$essure function.
	
In the following sections, various types of analysis like asymptotic analysis, linear perturbation and matrix stability analyses are carried out to theoretically demonstrate the accuracy and robustness of the  modified HLL-CPS schemes.
\subsection{Asymptotic analysis of the proposed HLL-CPS scheme}
Asymptotic analysis of the modified HLL-CPS scheme is carried out to determine the correlation between the pressure fluctuations and Mach number. From equation (\ref{thornberuv}), we obtain the following relations for all the velocity components

\begin{equation}
z=\bar{z}M_* \hspace{1cm}\bar{u}_{il}^*=\bar{u}_{il} \hspace{1cm} \bar{u}^*_K=\bar{u}_{il}\pm\dfrac{1}{2}\bar{z}M_*\Delta_{il}\bar{u} \hspace{1cm} \Delta_{il}\bar{u}^*=\bar{z}M_*\Delta_{il}\bar{u}
\end{equation}

The continuity equation remains unchanged with the reconstructed velocities. The x and y-direction momentum equation in the non-dimensional form shown in equation (\ref{hllcps-lowspeed-after-rotation}) can be written as 
\begin{equation}\label{x-momentum-hllcps-nd2}
\begin{split}
&|\bar{\Omega}|\dfrac{\partial(\bar{\rho}\bar{u})_i}{\partial{}\bar{t}}+ \sum_{l\epsilon\upsilon(i)}\left(\bar{u}_{n,il}\bar{\rho}_l(\bar{u}_{il}\pm\bar{z}M_*\Delta_{il}\bar{u})+\left(\dfrac{\bar{p}}{M_*^2}n_x\right)_{il}+\left(\dfrac{\bar{u}_{n}}{2\bar{a}}\dfrac{n_x}{M_*}\right)_{il}\Delta_{il}\bar{p}\right)\Delta{}s_{il} \\ &  -\sum_{l\epsilon\upsilon(i)}\left(\left(\dfrac{\bar{u}_{n}^2M_*}{2\bar{a}^3}-\dfrac{1}{2\bar{a}M_*}\right)_{il}\left(\bar{u}_{il}\Delta_{il}\bar{p}+(\bar{p}\bar{z}M_*)_{il}\Delta_{il}\bar{u}\right)\right)\Delta{}s_{il}=0
\end{split}
\end{equation}
\begin{equation}\label{y-momentum-hllcps-nd2}
\begin{split}
&|\bar{\Omega}|\dfrac{\partial(\bar{\rho}\bar{v})_i}{\partial{}\bar{t}}+ \sum_{l\epsilon\upsilon(i)}\left(\bar{u}_{n,il}\bar{\rho}_l(\bar{v}_{il}\pm\bar{z}M_*\Delta_{il}\bar{v})+\left(\dfrac{\bar{p}}{M_*^2}n_y\right)_{il}+\left(\dfrac{\bar{u}_{n}}{2\bar{a}}\dfrac{n_x}{M_*}\right)_{il}\Delta_{il}\bar{p}\right)\Delta{}s_{il} \\ &  -\sum_{l\epsilon\upsilon(i)}\left(\left(\dfrac{\bar{u}_{n}^2M_*}{2\bar{a}^3}-\dfrac{1}{2\bar{a}M_*}\right)_{il}\left(\bar{v}_{il}\Delta_{il}\bar{p}+(\bar{p}\bar{z}M_*)_{il}\Delta_{il}\bar{v}\right)\right)\Delta{}s_{il}=0
\end{split}
\end{equation}
It can be seen that the numerical dissipation of the convective flux due to the velocity jump is shifted to a higher order with Thornber-type low Mach correction. Now, expanding the terms asymptotically using equation (\ref{expansion}), and rearranging the terms in the equal power of $M*$, we obtain 
\begin{itemize}
	\item {Order of $M_*^0$}
\begin{itemize}
\item{continuity equation}
\begin{equation}
|\bar{\Omega}|\dfrac{\partial\bar{\rho}_{0i}}{\partial{}\bar{t}}+\sum_{l\epsilon\upsilon(i)}\left(\bar{u}_{n0,il}\bar{\rho}_{0K}+\dfrac{1}{2\bar{a}_{0,il}}\Delta_{il}\bar{p}_1\right)\Delta{}s_{il}=0
\end{equation}
\item{x-momentum equation}
\begin{equation}
\begin{split}
& |\bar{\Omega}|\dfrac{\partial(\bar{\rho}_0\bar{u_0})_i}{\partial{}\bar{t}}+ \sum_{l\epsilon\upsilon(i)}\left(\bar{\rho}_{0K}(\bar{u}_{n0}\bar{u}_0)_{il}+(\bar{p}_2n_x)_{il}\right)\Delta{}s_{il}+\\ &\left(\left(\dfrac{\bar{u}_{n0}n_x}{2\bar{a}_0}\right)_{il}\Delta_{il}\bar{p}_1+\dfrac{1}{2\bar{a}_{0,il}}(\bar{u}_{0,il}\Delta_{il}\bar{p}_1+\bar{u}_{1il}\Delta_{il}\bar{p}_0+\bar{z}_{il}\bar{p}_{0,il}\Delta_{il}\bar{u}_0)\right)\Delta{}s_{il}=0
\end{split}
\end{equation}
\item{y-momentum equation}
\begin{equation}
\begin{split}
& |\bar{\Omega}|\dfrac{\partial(\bar{\rho}_0\bar{v_0})_i}{\partial{}\bar{t}}+ \sum_{l\epsilon\upsilon(i)}\left(\bar{\rho}_{0K}(\bar{u}_{n0}\bar{v}_0)_{il}+(\bar{p}_2n_y)_{il}\right)\Delta{}s_{il}+\\ &\left(\left(\dfrac{\bar{u}_{n0}n_y}{2\bar{a}_0}\right)_{il}\Delta_{il}\bar{p}_1+\dfrac{1}{2\bar{a}_{0,il}}(\bar{v}_{0,il}\Delta_{il}\bar{p}_1+\bar{v}_{1il}\Delta_{il}\bar{p}_0+\bar{z}_{il}\bar{p}_{0,il}\Delta_{il}\bar{v}_0)\right)\Delta{}s_{il}=0
\end{split}
\end{equation}
\end{itemize}
\item{Order of $M_*^{-1}$}
\begin{itemize}
	\item {continuity equation}
	\begin{equation}
	\sum_{l\epsilon\upsilon(i)}\dfrac{1}{2\bar{a}_{0,il}}\Delta_{il}\bar{p}_0=0
	\end{equation}
	\item {x-momentum equation}
\begin{equation}
\sum_{l\epsilon\upsilon(i)}\left((\bar{p}_1n_x)_{il}+\left(\dfrac{\bar{u}_{n0}}{2\bar{a}_0}n_x\right)_{il}\Delta_{il}\bar{p}_0+\dfrac{\bar{u}_{0,il}}{2\bar{a}_{0,il}}\Delta_{il}\bar{p}_0\right)=0
\end{equation}
\item {y-momentum equation}
	\begin{equation}
\sum_{l\epsilon\upsilon(i)}\left((\bar{p}_1n_y)_{il}+\left(\dfrac{\bar{u}_{n0}}{2\bar{a}_0}n_y\right)_{il}\Delta_{il}\bar{p}_0+\dfrac{\bar{v}_{0,il}}{2\bar{a}_{0,il}}\Delta_{il}\bar{p}_0\right)=0
\end{equation}
\end{itemize}
From the continuity equation, we obtain $\sum_{l\epsilon\upsilon(i)}\Delta_{il}\bar{p}_0=0$. Hence, the x and y-momentum equations reduces to
\begin{equation}
\sum_{l\epsilon\upsilon(i)}(\bar{p}_1n_x)_{il}=0 ,
\hspace{2cm}
\sum_{l\epsilon\upsilon(i)}(\bar{p}_1n_y)_{il}=0
\end{equation}
The above equation  implies that 
\begin{equation}
\bar{p}_{1,i-1,j}-\bar{p}_{1,i+1,j}=0, \;\bar{p}_{1,i,j-1}-\bar{p}_{1,i,j+1}=0 
\end{equation}
This implies that $\bar{p}_1=$ constant for $\mathbf{i}$.
\end{itemize}

Thus, the modified HLL-CPS scheme supports pressure fluctuations of the type $p(x,t)=P_0(t)+M_*^2p_2(x,t)$ and hence shall be capable of resolving the low Mach flow features.

The leading order terms ( order of $M_*^0$ ) of the x and y-momentum equations are
\begin{equation}  \label{momentum-le-thornber}
\begin{array}{c}
|\bar{\Omega}|\dfrac{\partial(\bar{\rho}_0\bar{u_0})_i}{\partial{}\bar{t}}+\sum_{l\epsilon\upsilon(i)}\left(\bar{\rho}_{0K}(\bar{u}_{n0}\bar{u}_0)_{il}+(\bar{p}_2n_x)_{il}+\dfrac{1}{2}\left(\dfrac{\bar{z}\bar{p}_0}{\bar{a}_0}\right)_{il}\Delta_{il}\bar{u}_0\right)\Delta{}s_{il}=0 \\ \\ 
|\bar{\Omega}|\dfrac{\partial(\bar{\rho}_0\bar{v_0})_i}{\partial{}\bar{t}}+\sum_{l\epsilon\upsilon(i)}\left(\bar{\rho}_{0K}(\bar{u}_{n0}\bar{v}_0)_{il}+(\bar{p}_2n_y)_{il}+\dfrac{1}{2}\left(\dfrac{\bar{z}\bar{p}_0}{\bar{a}_0}\right)_{il}\Delta_{il}\bar{v}_0\right)\Delta{}s_{il}=0
\end{array}
\end{equation}
Comparing the above equations with that of the original scheme shown in equations (\ref{x-mom-om0-hllcps}, \ref{y-mom-om0-hllcps}), it can be seen that the numerical  dissipation of the leading order terms of the momentum equations is significantly reduced with the low Mach corrections. A similar reduction in numerical dissipation was achieved by Osswald et al. \cite{ossw} by applying low Mach corrections to the normal and tangential velocity jumps in the numerical dissipation terms of the Roe scheme. Further, it can be shown that the numerical dissipation of the convective fluxes of the HLL-CPS scheme with the present Thornber-type method is lower than that with Rieper's low Mach correction.
The leading order terms of the x and y-momentum equations, when the low Mach corrections are applied only to the velocity jumps in numerical dissipation terms, as proposed by Rieper \cite{rieper2}, are
\begin{equation} \label{momentum-le-rieper1}
\begin{array}{c}
|\bar{\Omega}|\dfrac{\partial(\bar{\rho}_0\bar{u_0})_i}{\partial{}\bar{t}}+\sum_{l\epsilon\upsilon(i)}\left(\bar{u}_{n0,il}(\bar{\rho}_0\bar{u}_0)_K+(\bar{p}_2n_x)_{il}+\dfrac{(\bar{z}\bar{p}_0)_{il}}{2\bar{a}_{0,il}}\Delta_{il}\bar{u}_0\right)\Delta{}s_{il}=0 \\ \\ 
|\bar{\Omega}|\dfrac{\partial(\bar{\rho}_0\bar{v_0})_i}{\partial{}\bar{t}}+\sum_{l\epsilon\upsilon(i)}\left(\bar{u}_{n0,il}(\bar{\rho}_0\bar{v}_0)_K+(\bar{p}_2n_x)_{il}+\dfrac{(\bar{z}\bar{p}_0)_{il}}{2\bar{a}_{0,il}}\Delta_{il}\bar{v}_0\right)\Delta{}s_{il}=0
\end{array}
\end{equation}
The upwind convective flux can be expressed as a sum of central difference terms and numerical dissipation and can be written as
\begin{equation} \label{momentum-le-rieper2}
\begin{array}{c}
|\bar{\Omega}|\dfrac{\partial(\bar{\rho}_0\bar{u_0})_i}{\partial{}\bar{t}}+\sum_{l\epsilon\upsilon(i)}\left(\bar{\rho}_{0K}\bar{u}_{n0,il}(\bar{u}_{0,il}\pm\Delta_{il}\bar{u}_0)+(\bar{p}_2n_x)_{il}+\dfrac{(\bar{z}\bar{p}_0)_{il}}{2\bar{a}_{0,il}}\Delta_{il}\bar{u}_0\right)\Delta{}s_{il}=0 \\ \\ 
|\bar{\Omega}|\dfrac{\partial(\bar{\rho}_0\bar{v_0})_i}{\partial{}\bar{t}}+\sum_{l\epsilon\upsilon(i)}\left(\bar{\rho}_{0K}\bar{u}_{n0,il}(\bar{v}_{0,il}\pm\Delta_{il}\bar{v}_0)+(\bar{p}_2n_y)_{il}+\dfrac{(\bar{z}\bar{p}_0)_{il}}{2\bar{a}_{0,il}}\Delta_{il}\bar{v}_0\right)\Delta{}s_{il}=0
\end{array}
\end{equation}
It can be seen from the above equation that the numerical dissipation of the upwind convective flux of the HLL-CPS scheme does not reduce with the low Mach correction of Rieper. Thus, we can conclude that, for the original HLL-CPS scheme, the numerical dissipation of the leading order terms of the momentum equation with the present Thornber-type method, shown in equation (\ref{momentum-le-thornber}) is lower than the numerical dissipation with Rieper's low Mach correction shown in equation (\ref{momentum-le-rieper2}).
\subsection{Linear perturbation analysis}
Linear perturbation analysis of the numerical scheme is performed for determining the stability of the scheme for high-speed flow problems. The evolution of the density, shear velocity, and pressure perturbations in the original and proposed HLL-CPS schemes are carried out like Quirk \cite{quirk} and Pandolfi-D'Ambrosio \cite{pand}. The normalized flow properties are described as 
\begin{equation}
\rho=1\pm \hat{\rho}, \hspace{1cm} u=u_0\pm \hat{u}, \hspace{1 cm} v=0, \hspace{1cm}  p=1\pm\hat{p}
\end{equation}
where $\hat{\rho}$, $\hat{u}$ and $\hat{p}$ are the density, shear velocity, and pressure perturbations. It may be noted that here $u$ is the shear velocity and $v$ is the face normal velocity.

The evolution of the perturbations in the proposed modified HLL-CPS-RTNP scheme is shown in Table \ref{pert-hllcps-new} along with other classical schemes of the HLL family. It can be seen from the table that  the density, shear velocity, and pressure perturbations are all damped in the HLLE scheme. It is observed that the pressure perturbations feed the density perturbations in the original HLL-CPS scheme, just like the classical Roe, HLLEM, and HLLC schemes. However, the shear velocity perturbations in the original HLL-CPS scheme are damped like the HLLE scheme. In the proposed HLL-CPS-RTNP scheme, the pressure function $f_p$ approaches unity  around a strong shock and hence the density and shear velocity perturbations are damped around a strong shock, just like the classical HLLE scheme. Hence, the proposed HLL-CPS-RTNP scheme is expected to be stable for strong shock. Across the isolated contact and shear discontinuities, the pressure function $f_p$ is zero and the density and shear velocity perturbation evolution in the HLL-CPS-RTNP scheme become identical to the classical Roe, HLLEM, and HLLC schemes.  Hence the proposed HLL-CPS-RTNP scheme will be able to resolve the isolated contact and shear discontinuities as crisply as the classical schemes.
\begin{table}[H]
\begin{center}
\caption{Result of Linear Perturbation Analysis of Improved HLL-CPS Schemes}
\label{pert-hllcps-new}
	\begin{tabular}{|c|c|c|c|c|}
	\hline Serial No & Scheme & $\hat{\rho}^{n+1}=$ & $\hat{u}^{n+1}=$  & $\hat{p}^{n+1}=$\\
	\hline 1 & HLLE & $\hat{\rho}^n(1-2\nu)$ &  $\hat{u}^n(1-2\nu)$ & $\hat{p}^n(1-2\nu)$\\
	\hline 2 & Original HLL-CPS & $\hat{\rho}^n-2\nu\dfrac{\hat{p}}{\gamma}$ &  $\hat{u}^n\left(1-\dfrac{2\nu}{\gamma}\right)$ & $\hat{p}^n(1-2\nu)$\\
	\hline 3 & Roe, HLLEM and HLLC & $\hat{\rho}^n-2\nu\dfrac{\hat{p}}{\gamma}$ &  $\hat{u}^n$ & $\hat{p}^n(1-2\nu)$\\
	\hline 4 & HLL-CPS-RTNP & $\hat{\rho}^n(1-2\nu{}f_p)-2\nu(1-f_p)\dfrac{\hat{p}}{\gamma}$& $\hat{u}^n\left(1-\dfrac{2\nu}{\gamma}f_p\right)$ & $\hat{p}^n(1-2\nu)$\\
	\hline
	\end{tabular}
\end{center}
\end{table}
\subsection{Matrix stability analysis}
In the present work, matrix stability analysis  is carried out like Dumbser et al. \cite{dumb-matrix}. A 2D computational domain $[0, 1] \times [0, 1]$ is considered and the domain is discretized into $11 \times 11$ grid points. The raw state is a steady normal shock wave with some perturbations
\begin{equation}
\mathbf{U}_\mathbf{i}=U^0_\mathbf{i}+\delta{}\mathbf{U}_\mathbf{i}
\end{equation}
where $\mathbf{i}=(i, j)$, $\mathbf{U}^0_\mathbf{i}$ is the solution of a steady shock and $\delta{}\mathbf{U}_\mathbf{i}$ is a small numerical random perturbation.
Substituting the expression into finite volume formulation, and after some evolution, the perturbation can be written as
\begin{equation}
\left(\begin{array}{c} \delta{}\mathbf{U}_\mathbf{1} \\ . \\ . \\ . \\ \delta{}\mathbf{U}_\mathbf{i}\end{array}\right)=exp^{\mathbf{S}t}\left(\begin{array}{c} \delta{}\mathbf{U}_\mathbf{1} \\ . \\ . \\ . \\ \delta{}\mathbf{U}_\mathbf{i}\end{array}\right)_0
\end{equation}
where $\mathbf{S}$ is the stability matrix based on the approximate Riemann solver. The perturbations will remain bounded if the maximum of the real part of the eigenvalues of the matrix $\mathbf{S}$ is non-positive, i.e.,
\begin{equation}
max(Re(\lambda(\mathbf{S})))\le 0
\end{equation}
The initial data is provided by the exact Rankine-Hugoniot solution in the x-direction. The upstream and downstream states are
\begin{equation}
\mathbf{U}_L=\left(\begin{array}{c} 1\\ 1\\ 0\\ \dfrac{1}{\gamma(\gamma-1)M_0^2}+\dfrac{1}{2}\end{array}\right) \text{for x}\le 0.5, \hspace{1cm} \mathbf{U}_R=\left(\begin{array}{c} f(M_0)\\ 1\\ 0\\ \dfrac{g(M_0)}{\gamma(\gamma-1)M_0^2}+\dfrac{1}{2f(M_0)}\end{array}\right) \text{for x} > 0.5
\end{equation}
where $f(M_0)=\left( \dfrac{2}{\gamma+1}\dfrac{1}{M^2_a}+\dfrac{\gamma-1}{\gamma+1}\right)^{-1}$,  $g(M_0)=\left( \dfrac{2\gamma}{\gamma+1}{M^2_a}-\dfrac{\gamma-1}{\gamma+1}\right)$ and  $M_0$ is the upstream Mach number.

\begin{figure}[H]
	\begin{center}
	\includegraphics[width=230pt]{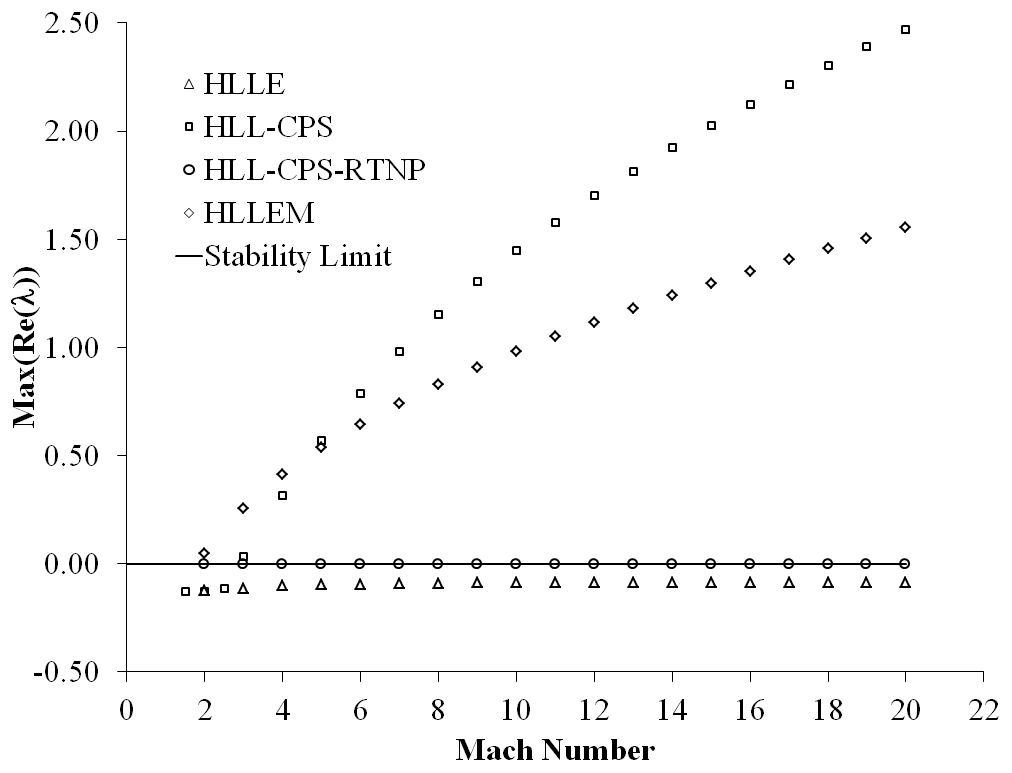}
	\caption{Plot of maximum real eigenvalues of $\mathbf{S}$ vs Mach number for the HLL-CPS and HLL-CPS-RTNP  Schemes. The results of the HLLE and HLLEM schemes are also shown for comparison.}
	\label{stability-plot}
	\end{center}
\end{figure}

A random perturbation of $10^{-6}$ is introduced to all the conserved variables of all the grid cells. The matrix stability analysis is carried out with a `thin' shock comprising only upstream and downstream values. The plot of the maximum real eigenvalues versus Mach number is shown in Fig. \ref{stability-plot}. It can be seen from the figure that the proposed HLL-CPS-RTNP scheme is stable like the HLLE scheme, while the original HLL-CPS and the HLLEM schemes are unstable.
In the next section, numerical results are presented for several carefully chosen test cases ranging from very low to very high Mach numbers to demonstrate the all Mach handling capability of the proposed HLL-CPS scheme. 

\section{Results}
Several test cases are presented to demonstrate numerically the properties of the proposed HLL-CPS-RTNP scheme and verify the results of theoretical analyses. First, a one-dimensional test case is presented to demonstrate the positively conservative property \cite{einf2} of the HLL-CPS schemes. The performance of the proposed HLL-CPS-RTNP scheme is then evaluated for high-speed flow problems to demonstrate the robustness of the scheme to numerical shock instabilities. For the high-speed test cases, first-order computations are carried out since the numerical instabilities are very prominent in the first-order schemes. Flow over a cylinder is solved to demonstrate the low Mach number flow resolution capability of the proposed scheme. Viscous results over a flat plate and a lid-driven cavity are also presented to demonstrate the enhanced shear layer and low Mach flow resolution capability of the proposed HLL-CPS-RTNP scheme.
\subsection{One-dimensional test case}
\begin{figure}[H]
\begin{center}
\includegraphics[width=200pt]{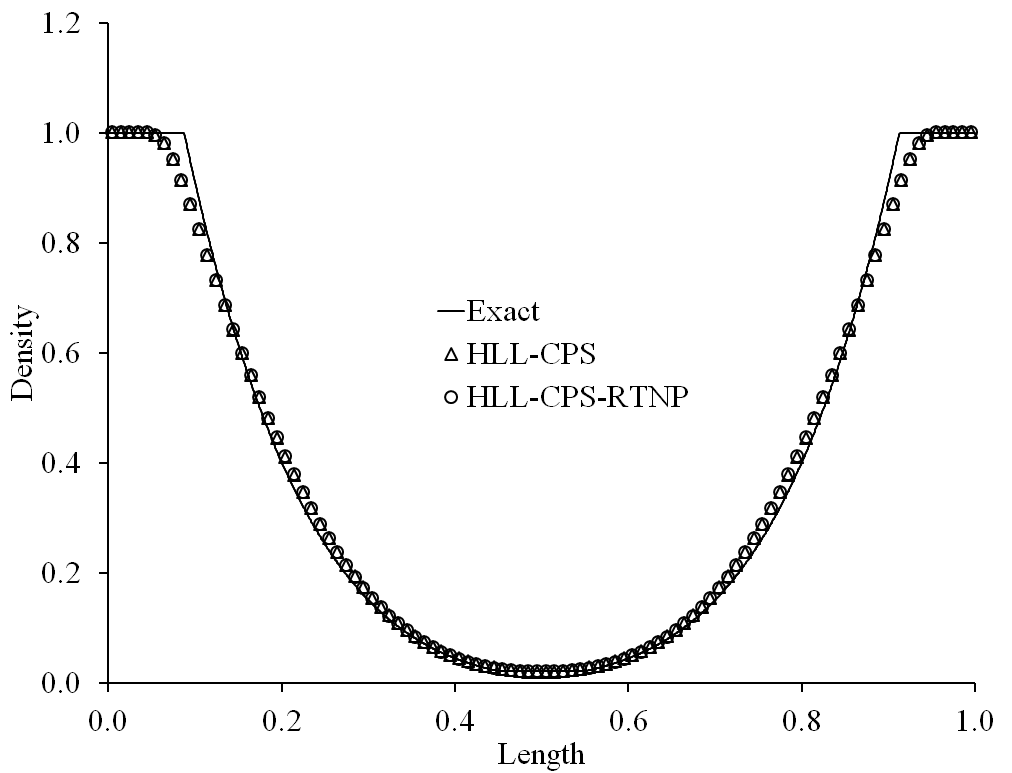}\includegraphics[width=200pt]{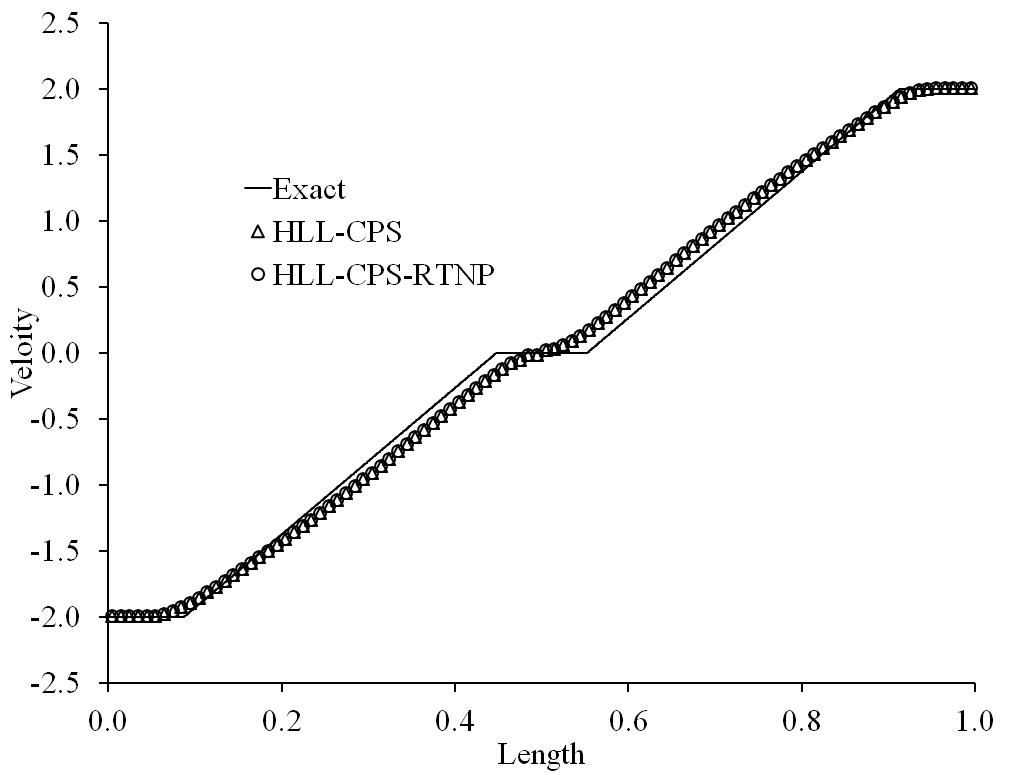}
\includegraphics[width=200pt]{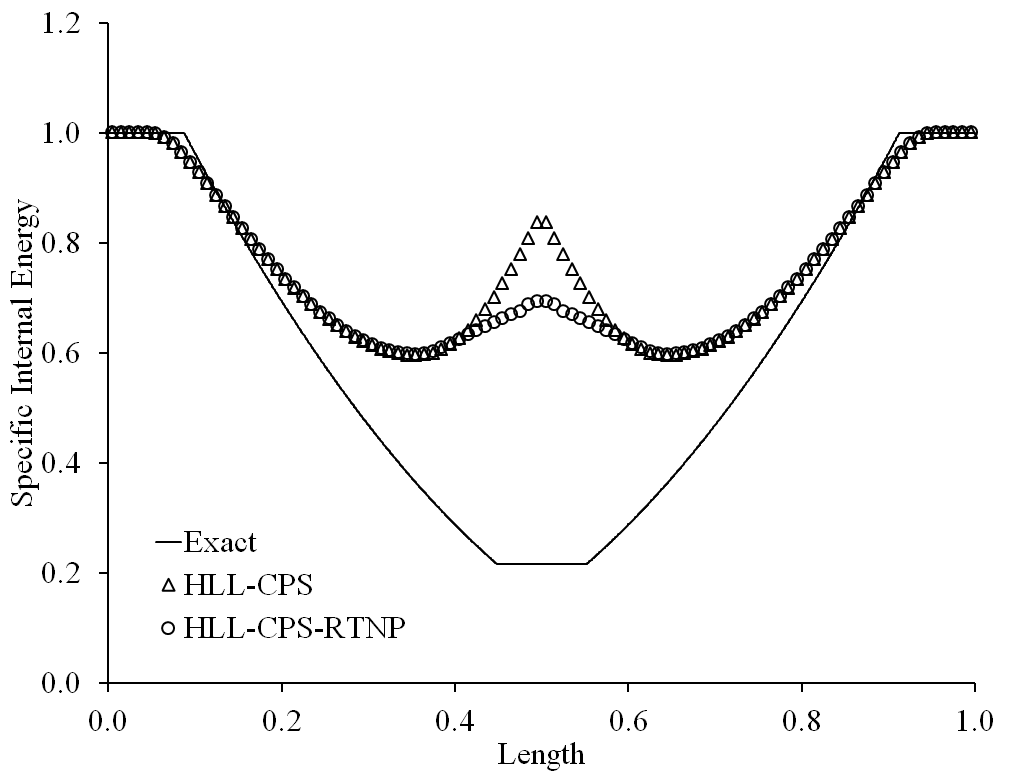}\includegraphics[width=200pt]{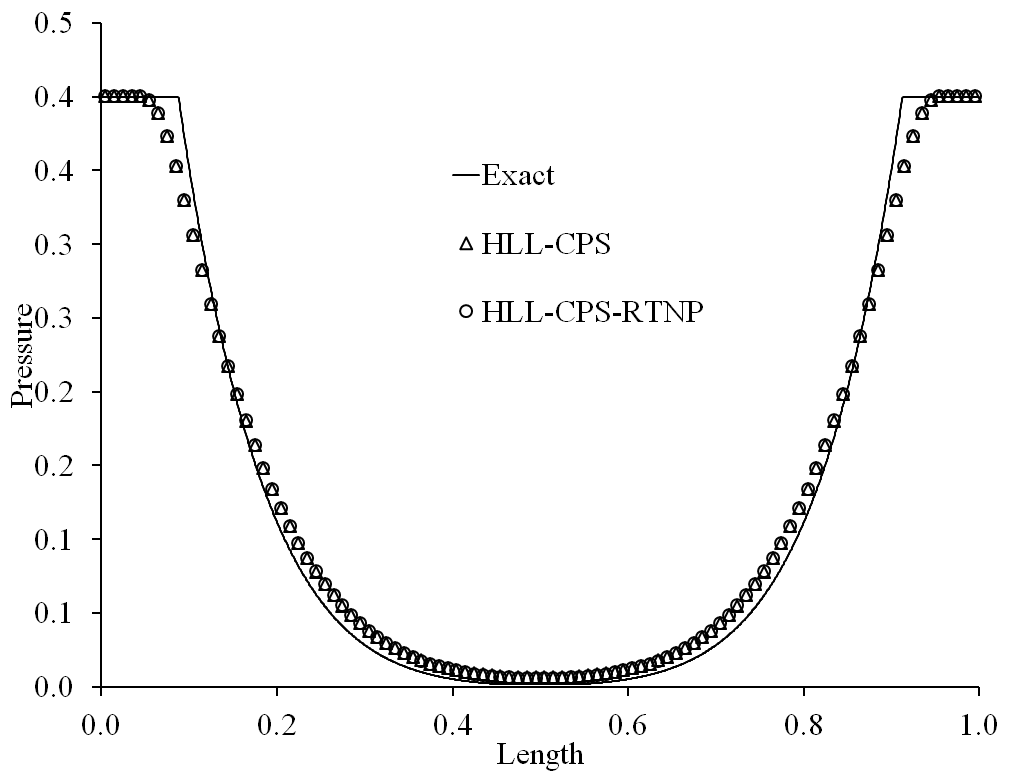}
\caption{Results of the HLL-CPS and HLL-CPS-RTNP schemes for the positively conservative test case.}
\label{test-case-2-hllcps}
\end{center}
\end{figure}

 In order to verify whether the proposed scheme is positively conservative \cite{einf2}, the scheme is tested on the Sod shock tube test case comprising two expansion fans running outwards on either side and a stationary contact. At time $t=0$, the values of ($\rho{}$, $u$, $p$) at the left and right of the interface located at $x=0.5$ are taken as (1.0, -2.0, 0.4) and (1.0, 2.0, 0.4) respectively. The computations are carried out on a grid of 101 points with a CFL number of 0.50. The results are shown in Fig. \ref{test-case-2-hllcps} at time $t=0.15$ sec. It can be seen from the figure that density and internal energy are positive even in the low-density region. The density and specific internal energy plots of the modified scheme are marginally closer to the analytical results than the original HLL-CPS scheme.
\subsection{Inviscid test cases}
 \subsubsection{Planar shock problem}
The problem consists of a Mach 6.0 shock wave propagating down in a rectangular channel \cite{quirk}. The domain consists of 800$\times{}$20 cells. The centreline in the y-direction (11$^{th}$ grid line) is perturbed to promote odd-even decoupling along the length of the shock as

\begin{figure}[H]
	\begin{center}
	\includegraphics[width=225pt]{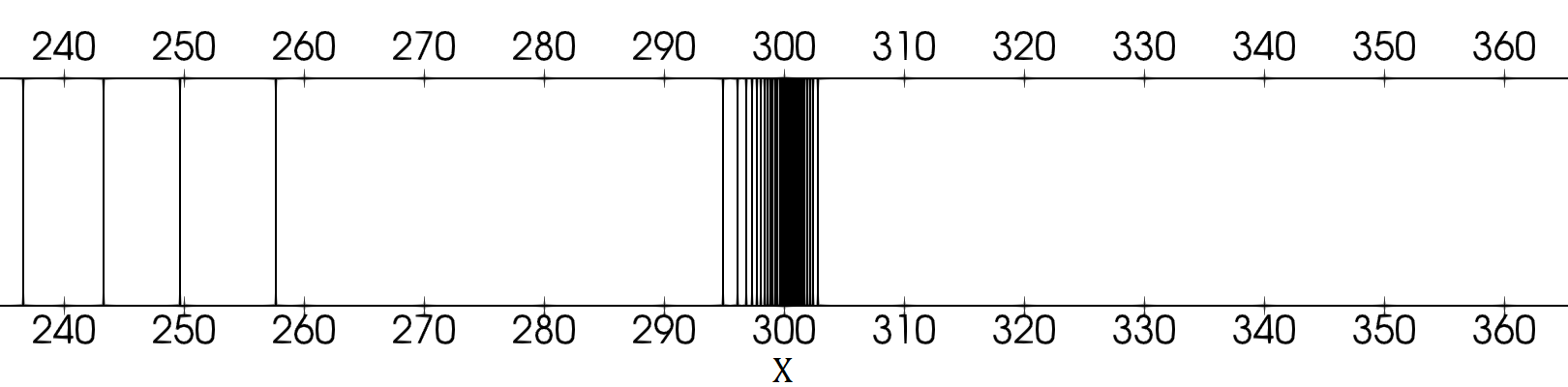} \includegraphics[width=225pt]{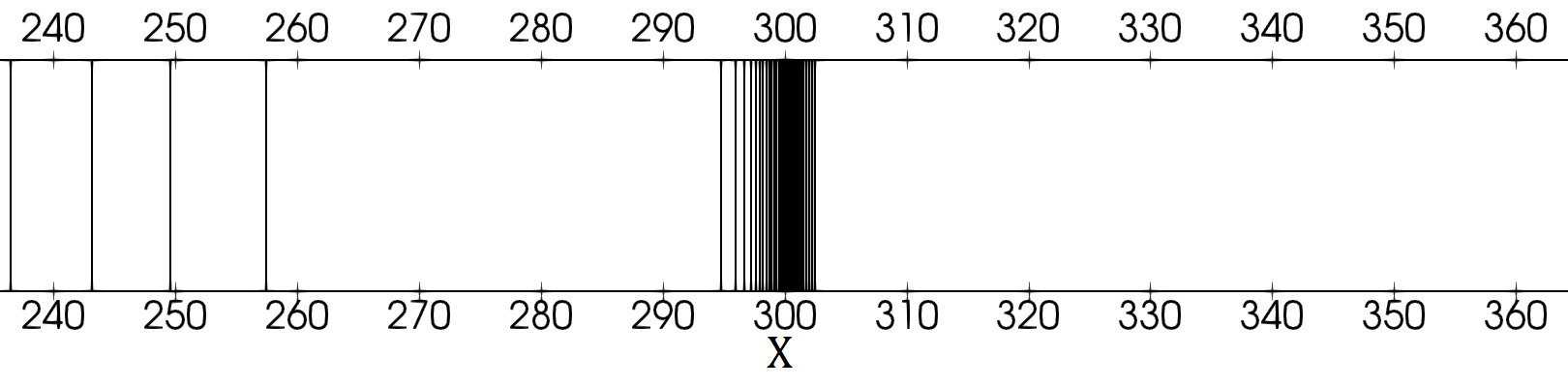}
	\\ (a) HLL-CPS Scheme \hspace{5cm} (b) HLL-CPS-RTNP Scheme \\
	\caption{Density contours for the Mach 6 planar shock problem}
	\label{planar-shock}
	\end{center}
\end{figure}

$y_{i,11}=\left\lbrace\begin{array}{c}\ y_{i, 11}+0.001 \hspace{3mm}\text {if i is even}\\ 
y_{i,11}-0.001 \hspace{3mm}\text {if i is odd} \end{array}\right.$. 
The test case was initially proposed by Quirk \cite{quirk} and later has been used by many scholars to demonstrate the odd-even decoupling instability. It has been reported that solvers that explicitly identify the contact would suffer from odd-even decoupling. The domain has been initialized with $\rho{}=1.4$, $p=1$, $u=0$, and $v=0$. The inlet boundary condition is set to post-shock values, the outlet boundary condition is set to zero gradients, the top boundary is a solid wall and the bottom boundary is also a solid wall. A CFL number of 0.5 has been used for all the schemes to compute the solutions. The contour plots for density are shown in Fig. \ref{planar-shock} at the time level $t=55$, and 30 contour levels varying from 1.6 to 7.0 are shown. It can be seen from Fig. \ref{planar-shock} that the HLL-CPS scheme and the proposed HLL-CPS-RTNP are free from the odd-even decoupling problem.
\subsubsection{Double Mach reflection problem}
\begin{figure}[H]
	\begin{center}
	\includegraphics[width=225pt]{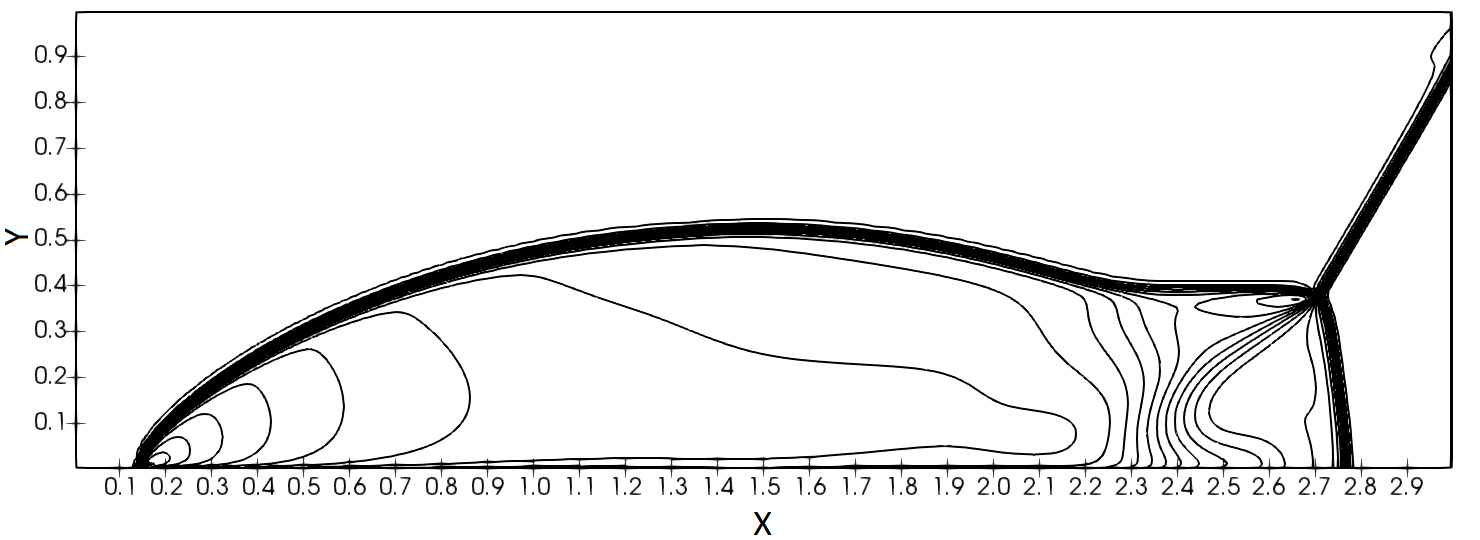} \includegraphics[width=225pt]{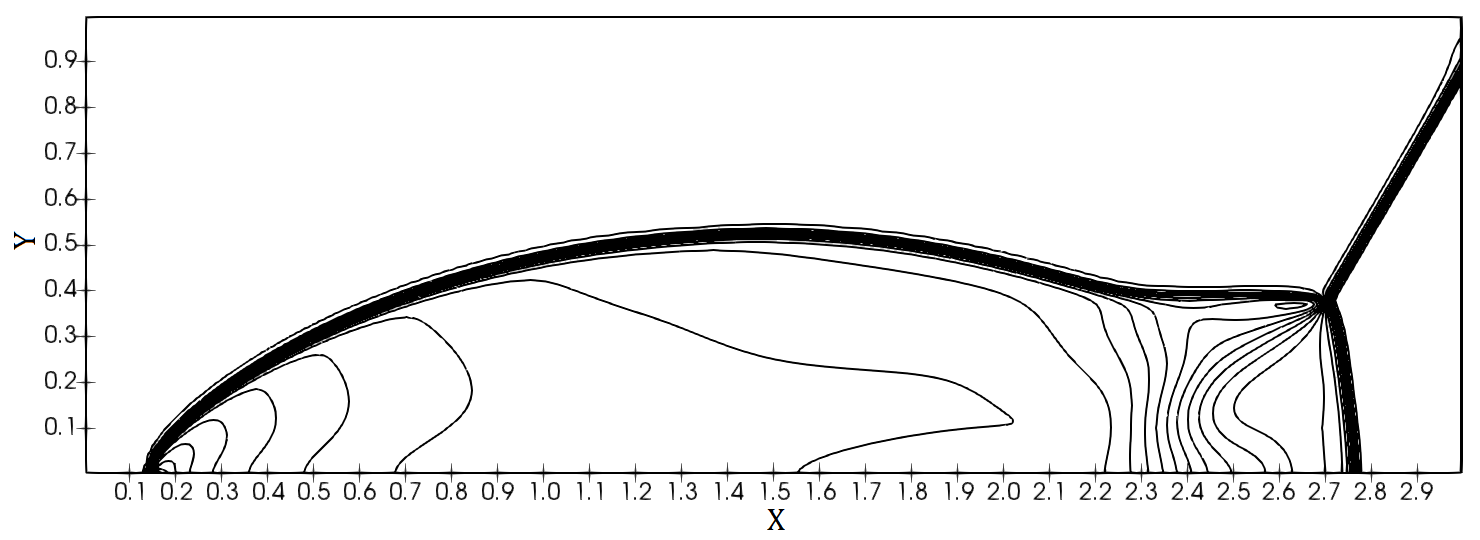}
	\\ (a) HLL-CPS Scheme \hspace{5cm} (b) HLL-CPS-RTNP Scheme \\
	\caption{Density contours for the double Mach reflection problem}
	\label{dmr}
	\end{center}
\end{figure}

The domain is four units long and one unit wide and is divided into 480$\times$ 120 cells. A Mach 10 shock initially making a 60-degree angle at x=1/6 with the bottom reflective wall is made to propagate through the domain. The domain ahead of the shock is initialized to the pre-shock values ($\rho=1.4$, $p=1$, $u=0$, $v=0$) and the domain behind the shock is assigned post-shock values. The inlet boundary condition is set to post-shock values and the outlet boundary is set to zero gradients. The top boundary is set to simulate the actual shock movement. At the bottom, the post-shock boundary condition is set up to x=1/6 and the reflective wall boundary condition is set thereafter. Density contours are shown in Fig. \ref{dmr} at time $t=2.00260\times 10^{-1}$. Thirty (30) contour levels ranging from 2.0 to 21.5 are shown in the figure. It can be seen from the figure that the HLL-CPS and HLL-CPS-RTNP schemes are free from the kinked Mach stem problem.
\subsubsection{Forward-facing step problem}
\begin{figure}[H]
	\begin{center}
	\includegraphics[width=225pt]{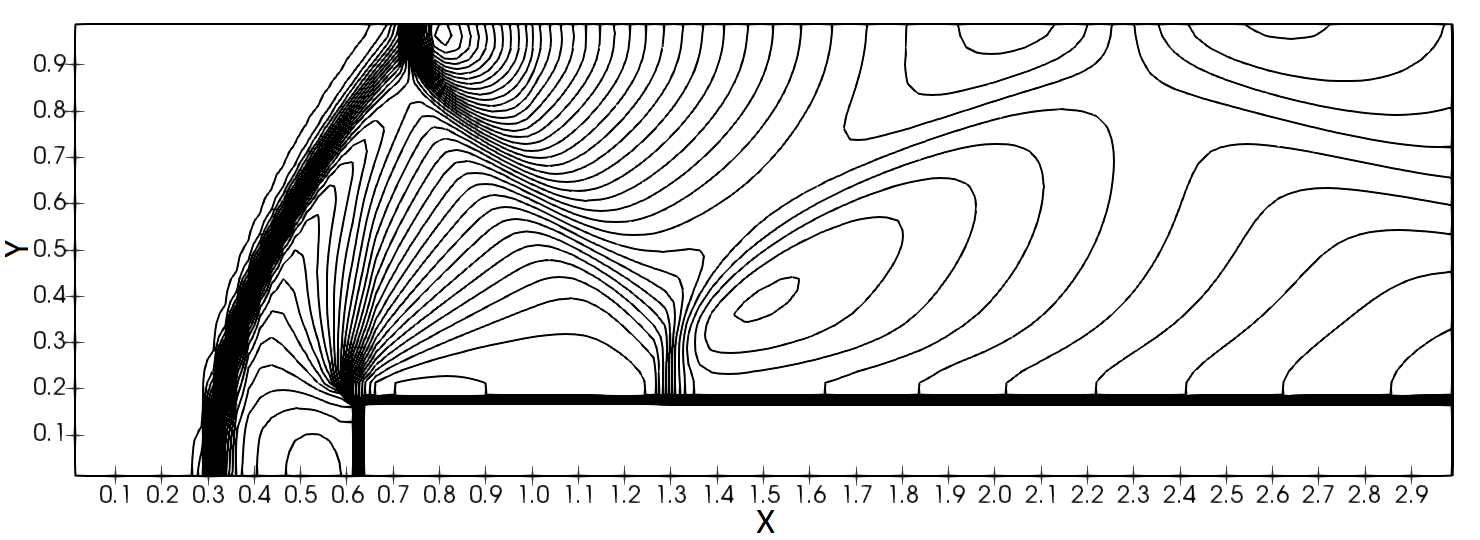} \includegraphics[width=225pt]{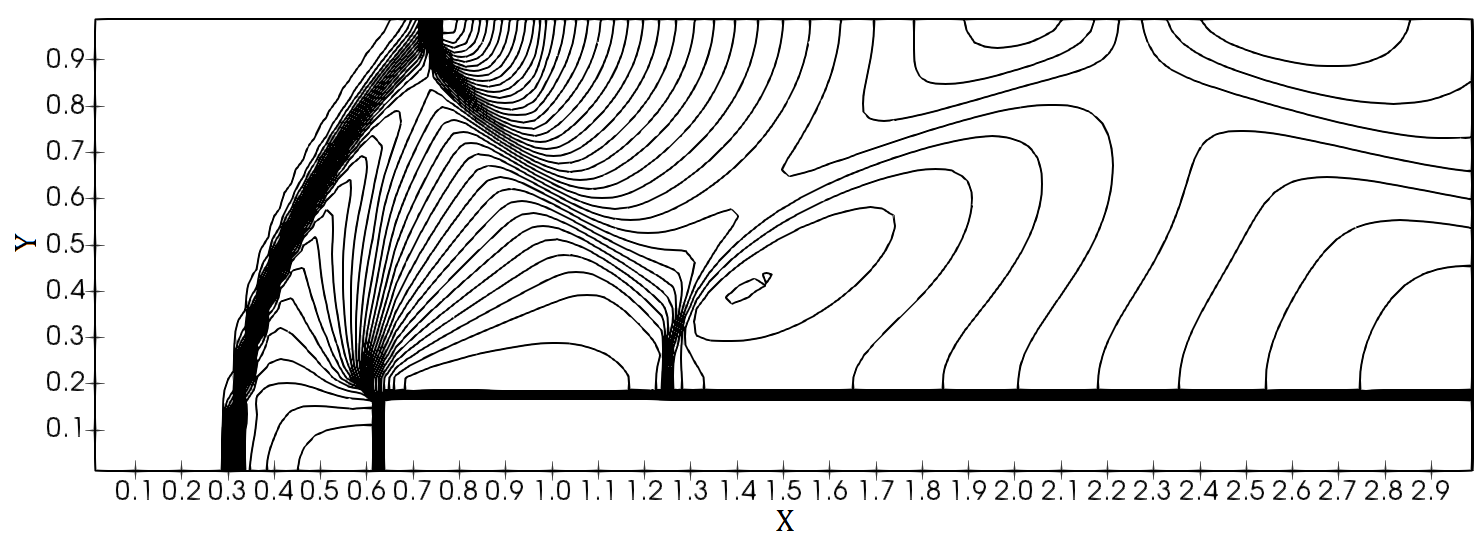}
	\\ (a) HLL-CPS Scheme \hspace{5cm} (b) HLL-CPS-RTNP Scheme \\
	\caption{Density contours for Mach 3 flow over a forward-facing step}
	\label{ffs}
	\end{center}
\end{figure}

The problem consists of a Mach 3 flow over a forward-facing step. The geometry consists of a step located 0.6 units downstream of the inlet and 0.2 units high. A mesh of $120 \times 40$ cells is used for a domain of 3 units long and 1 unit high. The complete domain is initialized with $\rho= 1.4$, $p=1$, $u=3$, and $v=0$. The inlet boundary is set to free-stream conditions, while the outlet boundary has zero gradients. At the top and bottom, reflective wall boundary conditions are set. Density contours are shown in Fig. \ref{ffs} at time $t=4.0$ units. A total of 45 contours from 0.2 to 7.0 are shown in the figure. It can be seen from the figure that the HLL-CPS and HLL-CPS-RTNP schemes capture the primary and reflected shocks without numerical instabilities. 

\subsubsection{Blunt body problem}
\begin{figure}[H]
	\begin{center}
		\includegraphics[width=90pt]{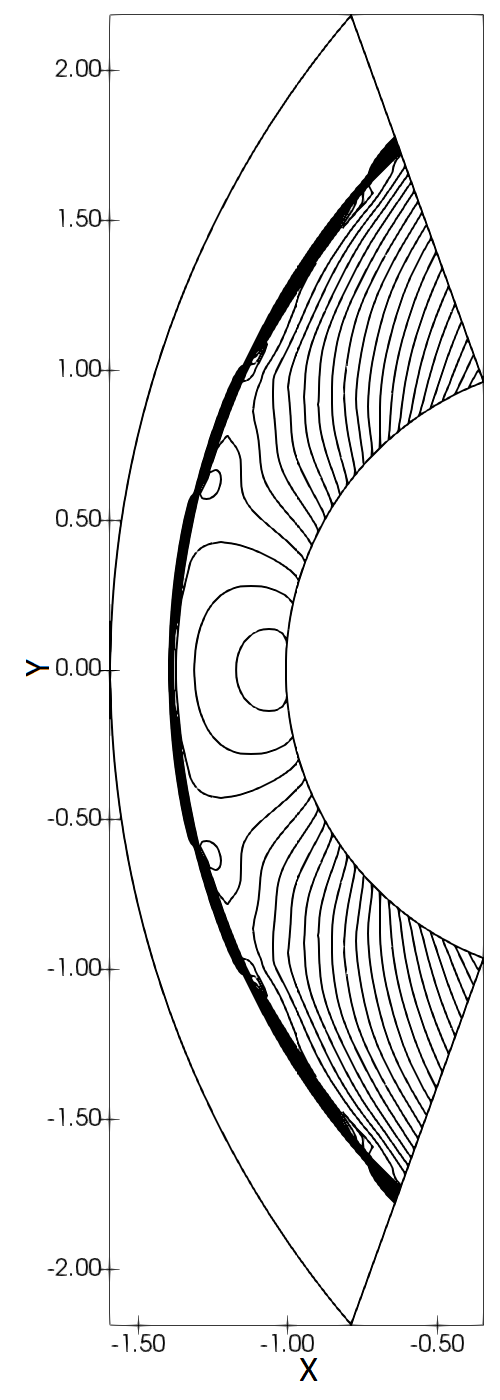}  \hspace{5mm}
		\includegraphics[width=90pt]{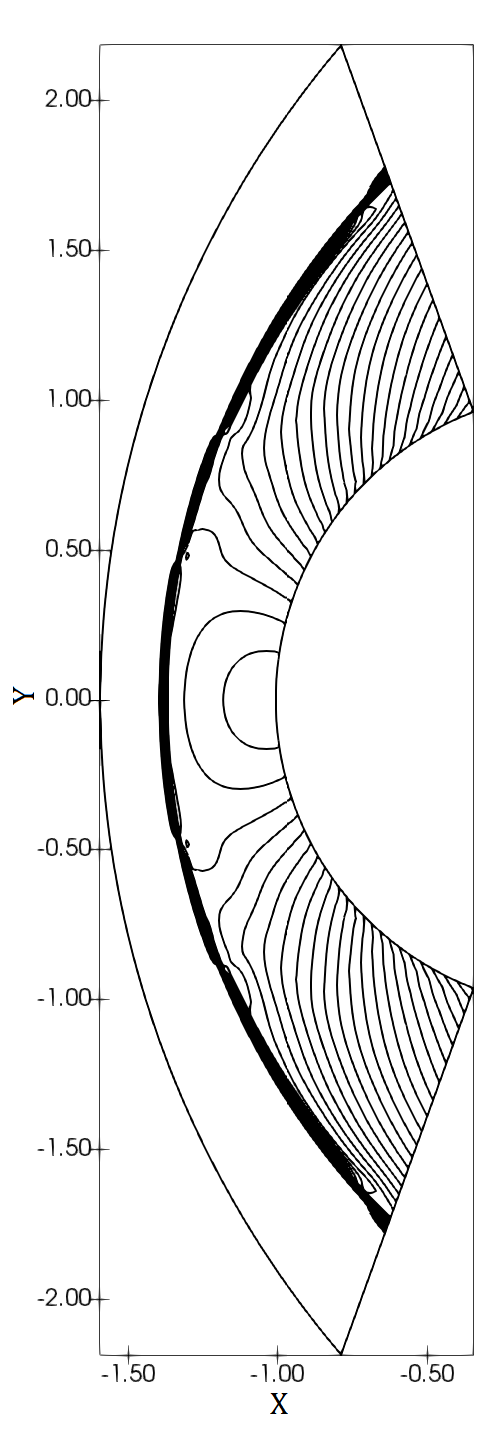} \\
		(a) HLL-CPS    \hspace{1cm}    (b)  HLL-CPS-RTNP 
		\caption{Density contour for Mach 20 flow over a blunt body}
		\label{carbuncle}
	\end{center}
\end{figure}

\begin{figure}[H]
	\begin{center}
	\includegraphics[width=200pt]{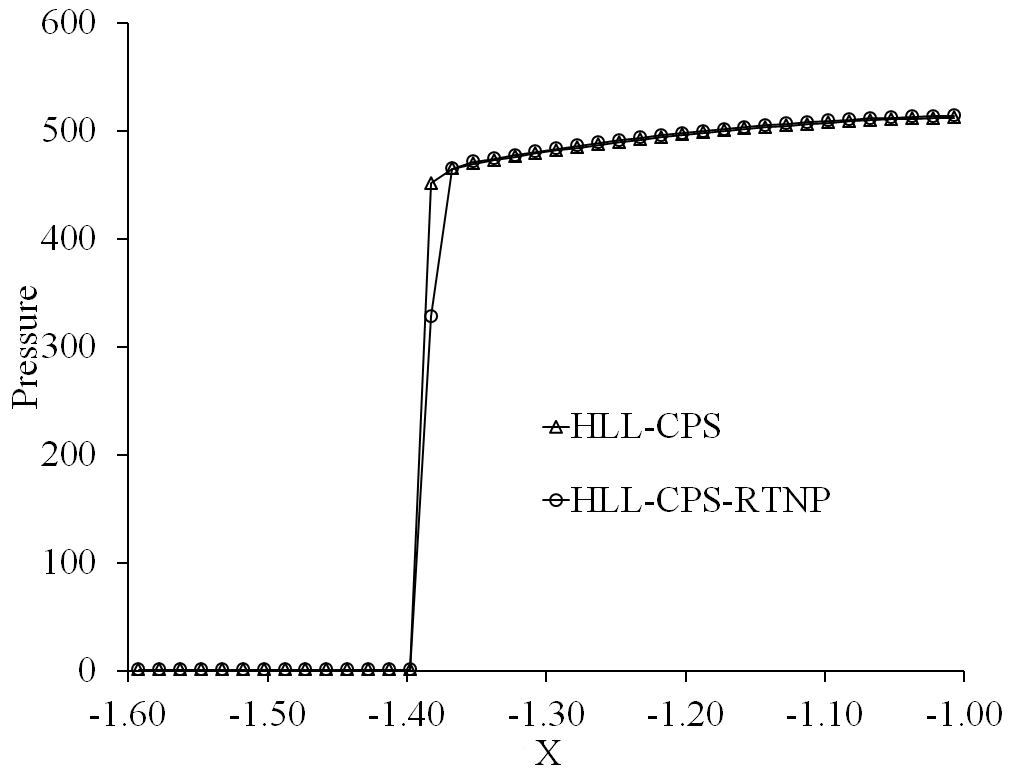} \\ (a) Overall \\
	\includegraphics[width=200pt]{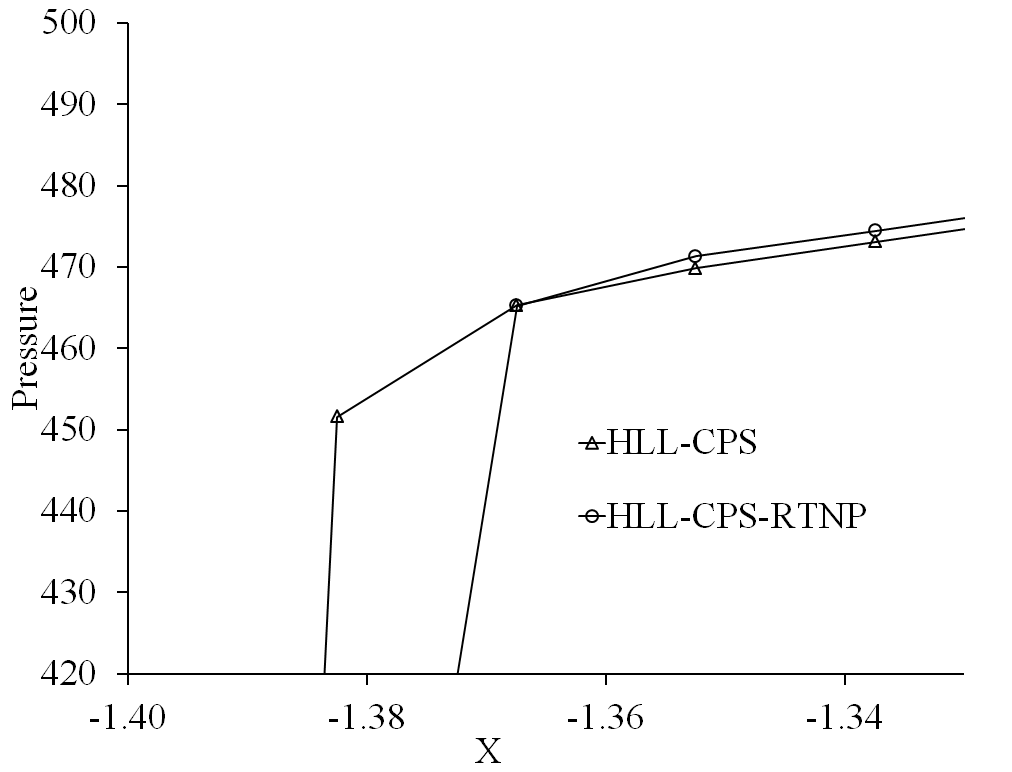}  
	\includegraphics[width=200pt]{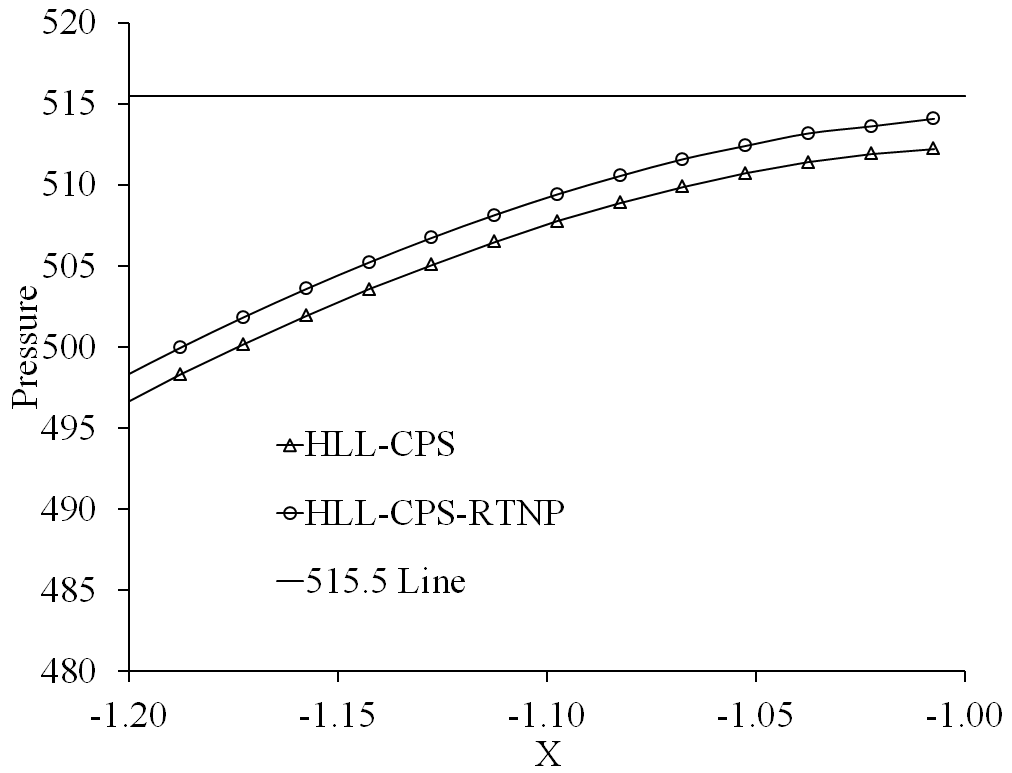} \\ (b) Post Shock \hspace{5cm} (c)  Near Stagnation Point
		\caption{Static pressure plot along the centreline ($y=0$) for the $M_{\infty}=20$ flow over a blunt body computed by the HLL-CPS scheme and HLL-CPS-RTNP schemes. The results are shown after 100,000 iterations}
	\label{centerline-carbuncle-hllcps}
	\end{center}
\end{figure}

\begin{figure}[H]
	\begin{center}
		\includegraphics[width=250pt]{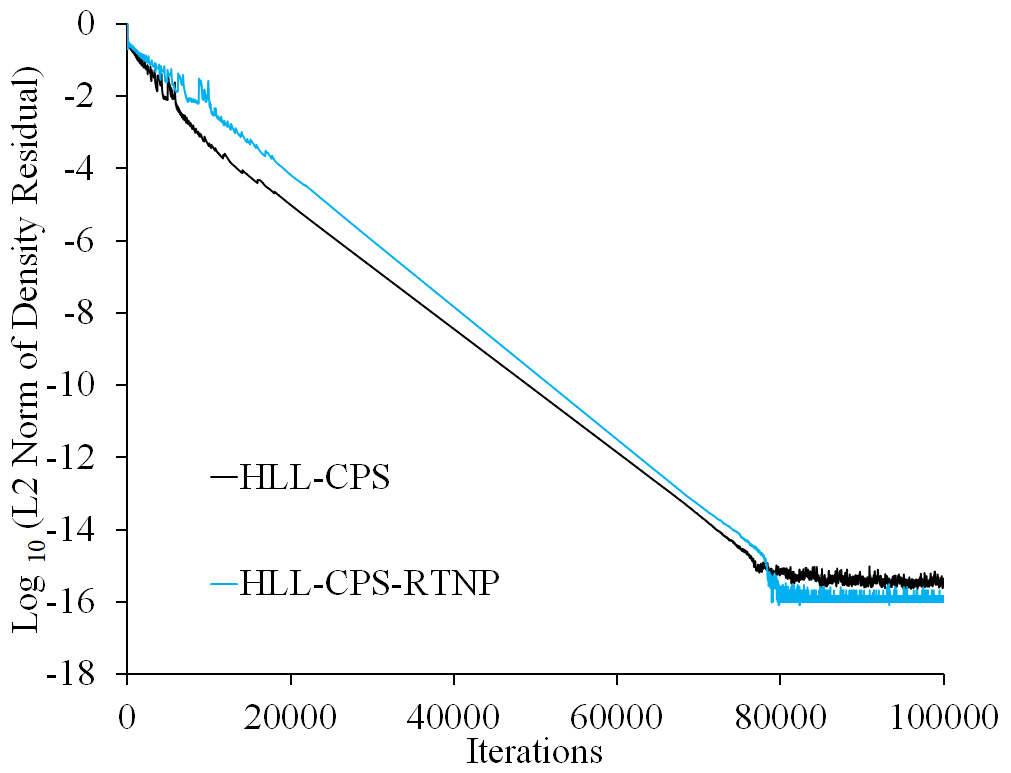} 
		\caption{Convergence history of the L-2 norm of density residual for the $M_{\infty}=20$ flow over a blunt body computed by the HLL-CPS and HLL-CPS-RTNP schemes.}
		\label{convergence-carbuncle-hllcps}
	\end{center}
\end{figure}

Hypersonic flow over a blunt body is a typical problem to assess the performance of a scheme for carbuncle instability. Free-stream Mach number of 20 is used in the computation. The grid for the blunt body had a size of 40 $\times{}$ 320 cells. The domain is initialized with values of $\rho=1.4$, $p=1$, $u=20$, and $v=0$. The inlet boundary condition is set to free-stream values and the solid wall boundary condition is applied to the blunt body. The density contour plots for the two schemes are shown in Fig. \ref{carbuncle}, and a total of 27 density contours from 2.0 to 8.7 are drawn. It can be seen from the figure that both the HLL-CPS and HLL-CPS-RTNP schemes are free from the carbuncle phenomenon. The static pressure plot along the centreline $(y=0)$ is shown in Fig. \ref{centerline-carbuncle-hllcps}. It can be seen from the figure that the original HLL-CPS scheme and the proposed HLL-CPS-RTNP scheme produce monotone solutions without any overshoot or undershoot. 
The post-shock and stagnation static pressure values are shown in Table \ref{hllcps-pressure}. It can be seen from the table that the post-shock static pressure of the original HLL-CPS scheme and the proposed HLL-CPS-RTNP scheme are very close to the analytical post-shock static pressure value of 466.5. It is also observed that the stagnation point static pressure of the proposed scheme with velocity reconstruction is closer to the analytical stagnation pressure value than the original HLL-CPS scheme. Therefore, it is felt that the low Mach corrections to the approximate Riemann solvers can contribute to more accurate solutions in the post-shock region of high-speed flows where the Mach number is subsonic. For this test case, the post-shock Mach number is about 0.38.
\begin{table}[H]
\caption {Static Pressure Comparison of HLL-CPS schemes for $M_{\infty}=20$ flow over a blunt body}
\label{hllcps-pressure}
\begin{center}
\begin{tabular}{|c|c|c|c|}
\hline Serial No & Scheme & Post-Shock Pressure (Pa)& Stagnation Pressure (Pa) \\ 
\hline  1 & Analytical & 466.5 & 515.5\\ 
\hline  2 & HLL-CPS & 465.2722 & 512.2300\\   
\hline  3 & HLL-CPS-RTNP & 465.2645 & 513.1969\\ 
\hline
\end{tabular}
\end{center}
\end{table}
The convergence history plots of the various HLL-CPS schemes are shown in Fig. \ref{convergence-carbuncle-hllcps}. It can be seen from the figure that the original HLL-CPS scheme and the proposed HLL-CPS-RTNP scheme converge to machine accuracy  within 100,000 iterations. 
\subsubsection{Supersonic corner problem}
\begin{figure}[H]
	\begin{center}
		\includegraphics[width=200pt]{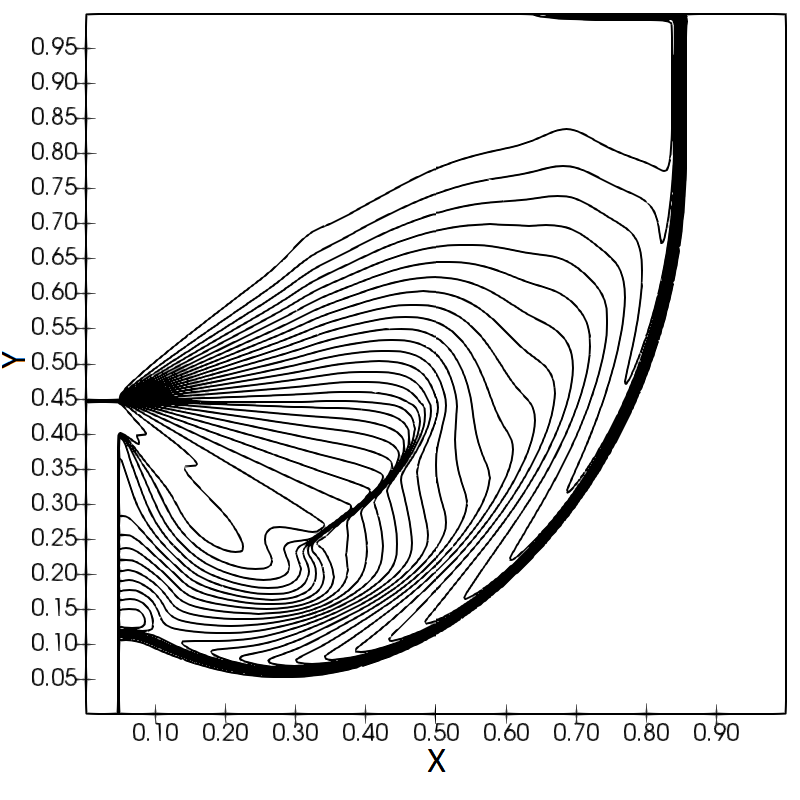} \hspace{5 mm} \includegraphics[width=200pt]{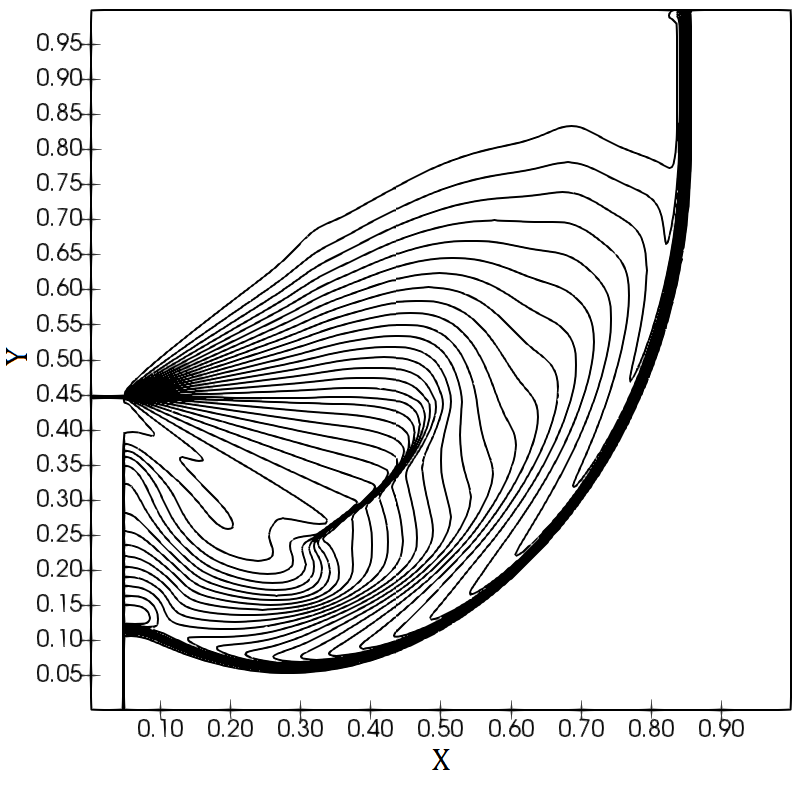} \\  
		(a) HLL-CPS Scheme  \hspace{3cm} (b) HLL-CPS-RTNP Scheme\\	
		\caption{Density contours for diffraction of Mach 5.09 normal shock around a 90$^{o}$ corner}
		\label{scr}
	\end{center}
\end{figure}

The problem consists of a sudden expansion of Mach 5.09 normal shock around a 90-degree corner. The domain is a square of one unit and is divided into $400\times400$ cells. The corner is located at $x=0.05$ and $y=0.45$. The initial normal shock is located at $x=0.05$. The domain to the right of the shock is assigned initially with pre-shock conditions of $\rho=1.4$, $p=1.0$, $u=0.0$, and $v=0.0$. The domain to the left of the shock is assigned post-shock conditions. The inlet boundary is supersonic, the outlet boundary has zero gradients and the bottom boundary behind the corner uses extrapolated values. Reflective wall boundary conditions are imposed on the corner. A CFL value of 0.8 is used and the density plots are generated for time $t=0.1561$ units. A total of 30 density contours ranging from 0 to 7.1 are shown in Fig. \ref{scr}. A benign perturbation can be seen in the HLL-CPS scheme around the shock location at the top boundary. On the other hand, the proposed HLL-CPS-RTNP scheme is free from the numerical shock instability problem of the original HLL-CPS scheme.  
\subsubsection{Flow around a cylinder at low Mach numbers}
\begin{figure}[H]
	\begin{center}
		\includegraphics[width=150pt]{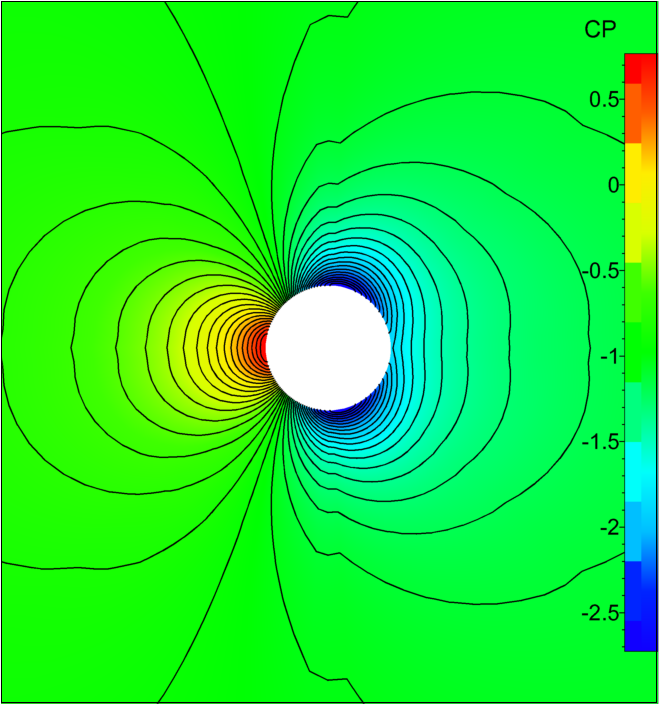} 
		\includegraphics[width=150pt]{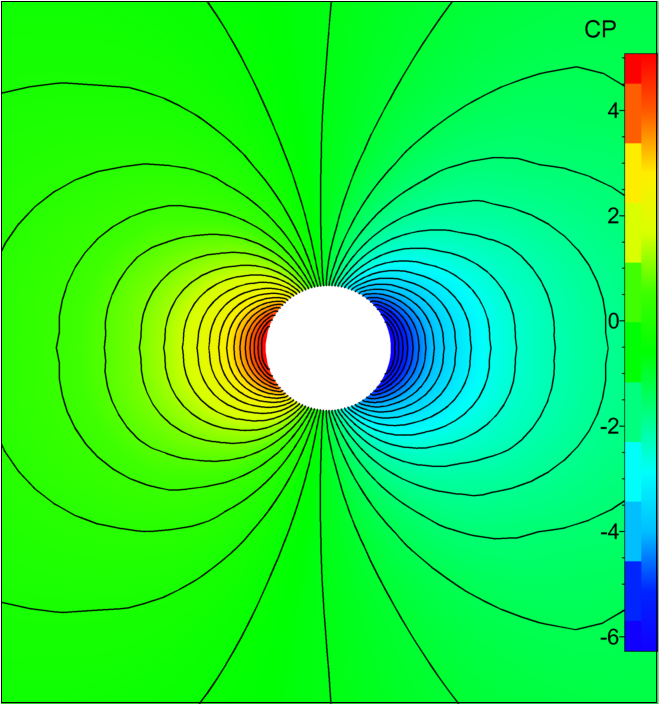} 
		\includegraphics[width=150pt]{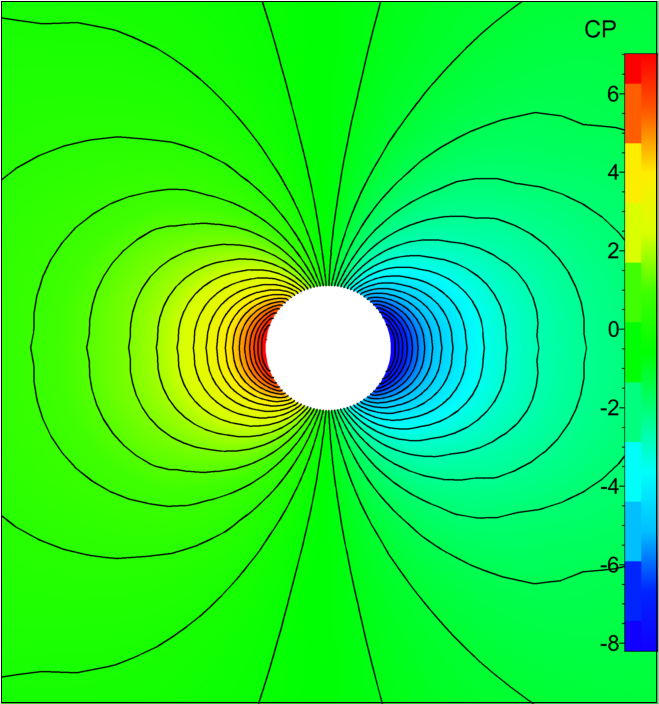}\\
		(a) Mach 0.10 \hspace{3cm} (b) Mach 0.01 \hspace{3cm}	(c) Mach 0.001 
		\caption{Static pressure coefficient contour plots for flow around a cylinder computed by the HLL-CPS scheme for free-stream Mach numbers of 0.10, 0.01 and 0.001}
		\label{cylinder-lm-hllcps}
	\end{center}
\end{figure}

\begin{figure}[H]
	\begin{center}	
		\includegraphics[width=150pt]{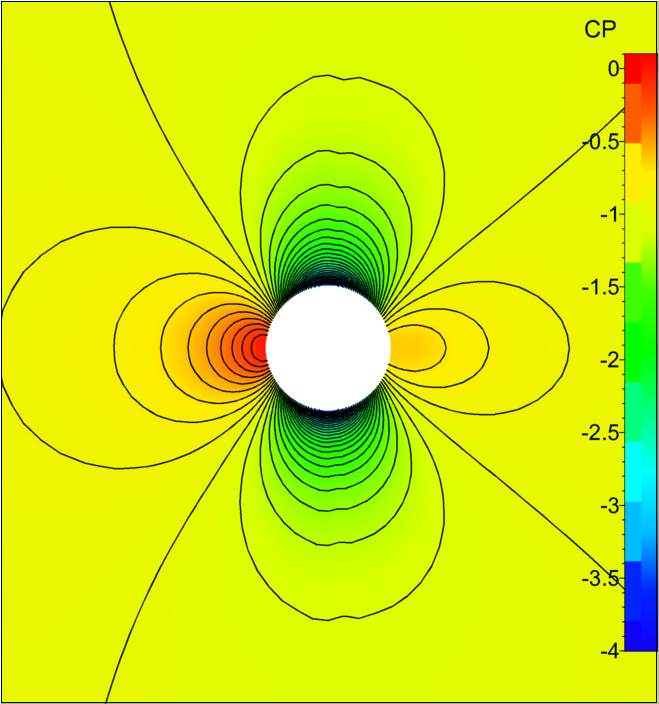}  
		\includegraphics[width=150pt]{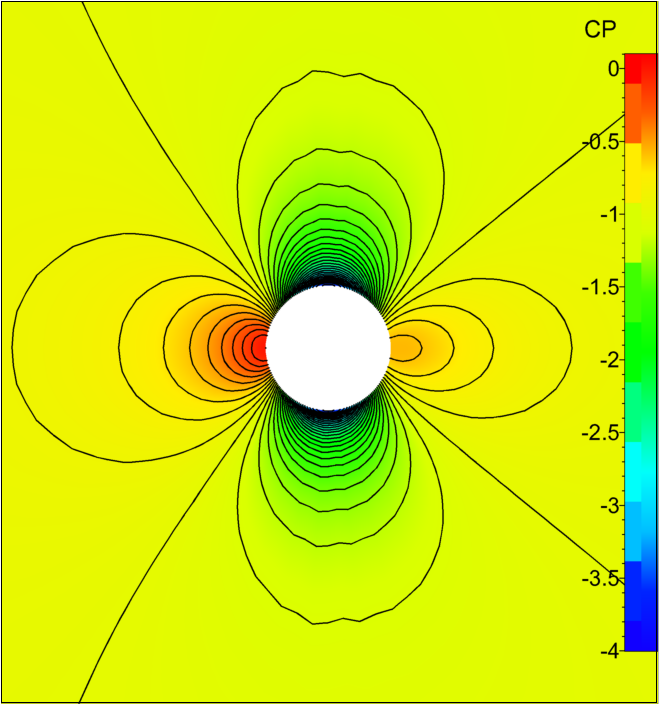}  
		\includegraphics[width=150pt]{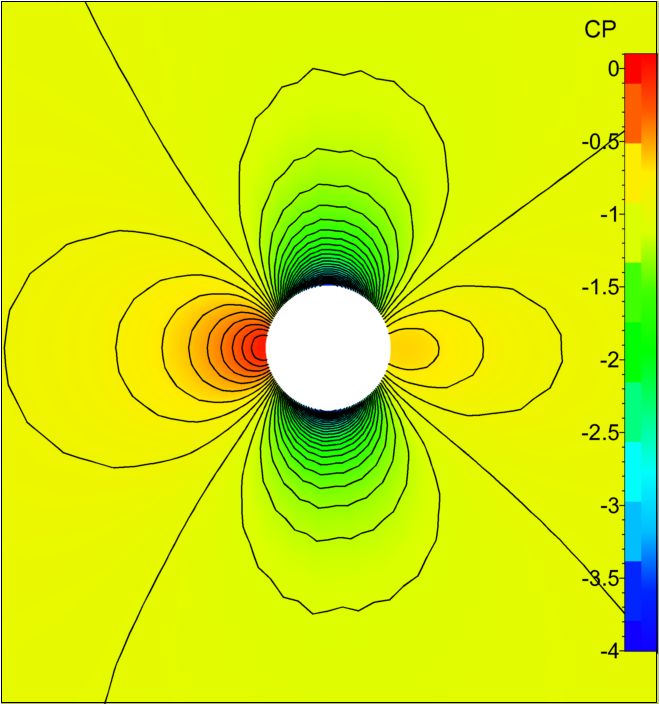} \\
		(a) Mach 0.100 \hspace{3cm} (b) Mach 0.010 \hspace{3cm} (c) Mach 0.001 \\
		\caption{Static pressure coefficient contour plots for flow around a cylinder computed by the HLL-CPS-RTNP scheme for free-stream Mach numbers of 0.10, 0.01 and 0.001}
		\label{cylinder-lm-hllcps-new}
	\end{center}
\end{figure}

Inviscid computations are carried out on a cylinder using the original HLL-CPS scheme and the proposed scheme at very low Mach numbers. Three grids of size $49\times19\times5$, $49\times37\times5$, and $97\times73\times5$ are used for the computations. The static pressure coefficient plots of the first-order HLL-CPS scheme for free-stream Mach numbers of 0.10, 0.01, and 0.001 are shown in Fig. \ref{cylinder-lm-hllcps} for the grid $97\times73\times5$. The static pressure coefficient is defined as $c_p=(p-p_{\infty})/(\frac{1}{2}\rho_{\infty}u_{\infty}^2)$ where $p_{\infty}$, $\rho_{\infty}$ and $u_{\infty}$ are the free-stream pressure, density and velocity respectively. Due to higher numerical viscosity in the HLL-CPS scheme, the pressure contours resemble creeping Stokes flow. The static pressure coefficient plot also shows the highest numerical viscosity for the Mach 0.001 case, indicating that numerical viscosity increases with a decrease in Mach number for the HLL-CPS scheme. The static pressure coefficient plots for the first-order HLL-CPS-RTNP scheme for Mach 0.1, 0.01, and 0.001 are shown in Fig. \ref{cylinder-lm-hllcps-new}. A total of 41 contours from -4.0 to 0.0 with intervals of 0.1 are drawn.  Due to lower numerical viscosity, the pressure plots for the proposed scheme resemble potential flow for the three Mach numbers cases. For the free-stream Mach number of 0.001, although a very weak checkerboard is seen for the proposed scheme, the accuracy of the results appears to be reasonable. For the second-order computations, the results of the proposed HLL-CPS-RTNP scheme resemble potential flow for all three Mach numbers even on a grid of $49\times19\times5$.
\subsection{ Viscous test cases}
\subsubsection{Flat plate boundary layer profile}
\begin{figure}[H]
\begin{center}
	\includegraphics[width=225pt]{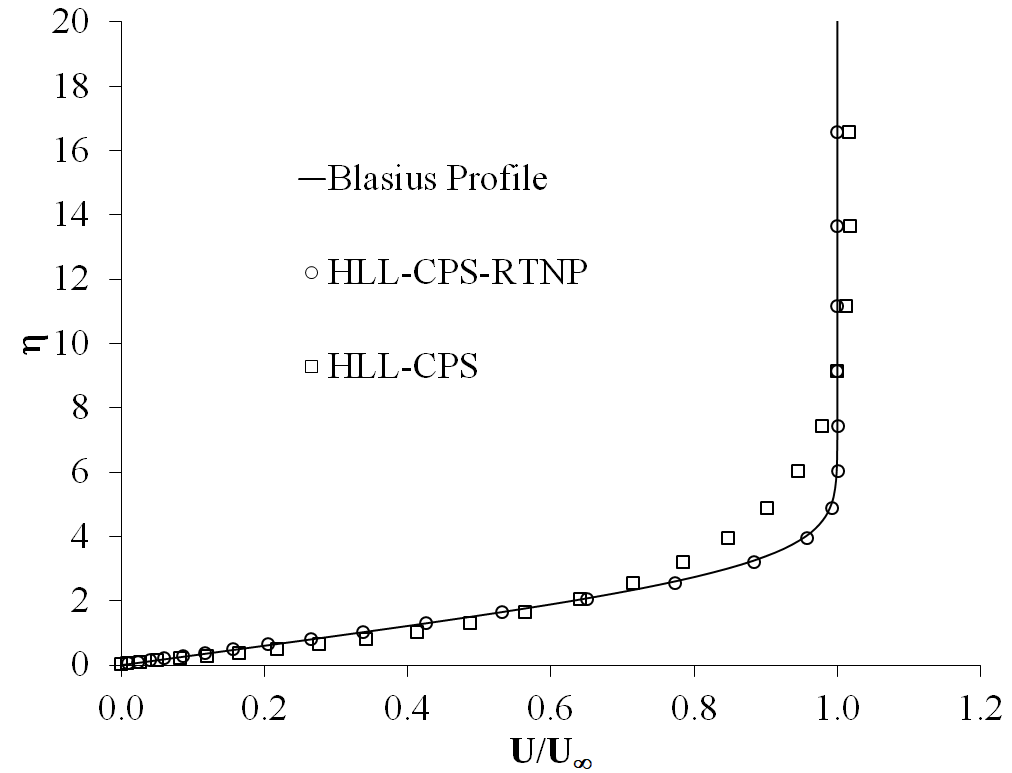}
	\caption{Boundary Layer profile for the $M_{\infty}=0.20$ laminar flow over a flat plate computed by HLL-CPS and HLL-CPS-RTNP schemes.}
	\label{bl}
\end{center}
\end{figure}

Computations are carried out over a flat plate with the proposed HLL-CPS-RTNP scheme along with the original scheme to demonstrate the boundary layer resolving capability of the proposed scheme. The computations are carried out for a free-stream  Mach number of 0.20 and Reynolds number of about 25,000. The third-order computations are carried out with the MUSCL approach \cite{leer} without any limiters. The boundary layer profiles are shown in Fig. \ref{bl} and the normalized velocity $\left(\frac{U}{U_{\infty}}\right)$ is plotted against the Blasius parameter $\eta=\sqrt{(U/\nu{}x)}$. The results are obtained after 50,000 iterations with a CFL number of 0.50 on a grid size of $81\times33\times5$. It can be seen that the HLL-CPS scheme is unable to resolve the boundary layer while the proposed HLL-CPS-RTNP scheme resolves the boundary layer accurately. The results demonstrate the effectiveness of the velocity reconstruction based on the face normal Mach number in the HLL-CPS scheme.
\subsubsection{Lid-driven cavity flow}
The lid-driven cavity is a benchmark test case for incompressible flow. The computations are carried out for  Reynolds numbers 100, 400, and 1000. Uniform grids of size 33$\times{}$33, 65$\times{}$65, and 129$\times{}$ 129 are used in the computations to study the grid dependency of the schemes. Second-order computations are carried out with the MUSCL approach \cite{leer} using the van Albada limiter \cite{albada}. However, it has been mentioned in \cite{thorn2} that all limiters could not be used with this low Mach correction for the HLLC scheme. Therefore, the van Leer \cite{leer} and the min-mod \cite{roe-review} limiters are also used for the case with a Reynolds number of 400 to demonstrate the compatibility of the low Mach corrections for the HLL-CPS scheme with different limiters. The velocity of the driven wall was 20 m/sec corresponding to a Mach number of about 0.05. The equations are solved by a two-stage TVD Runge-Kutta method of Gottlieb and Shu \cite{gott}. A CFL number of 0.50 is used for all the computations.

\begin{figure}[H]
	\includegraphics[width=200pt]{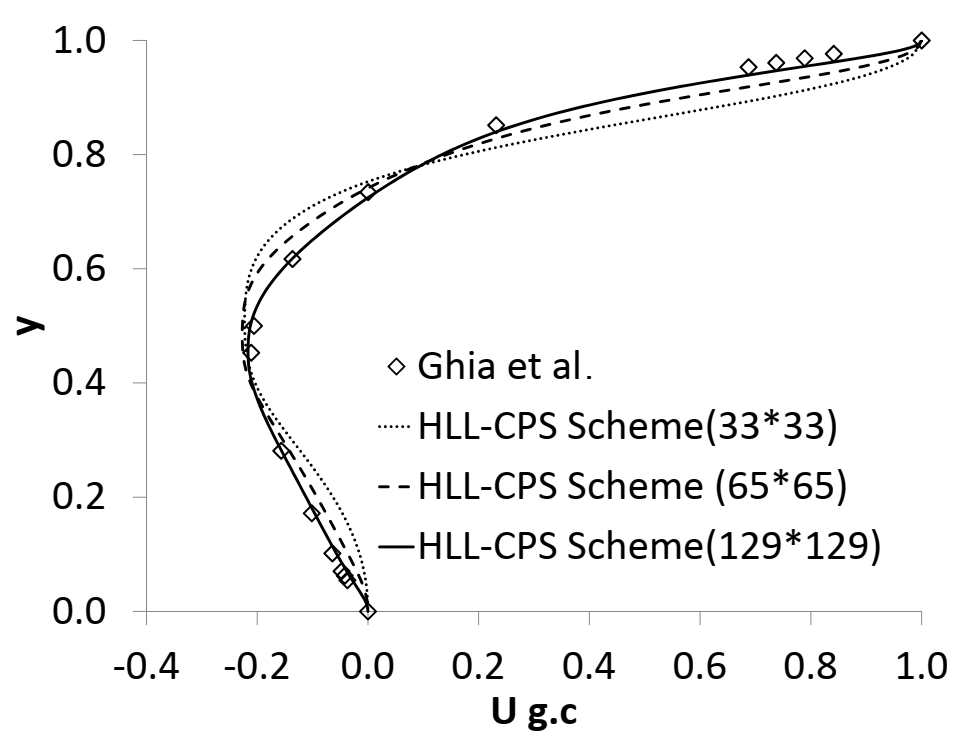}
	\includegraphics[width=200pt]{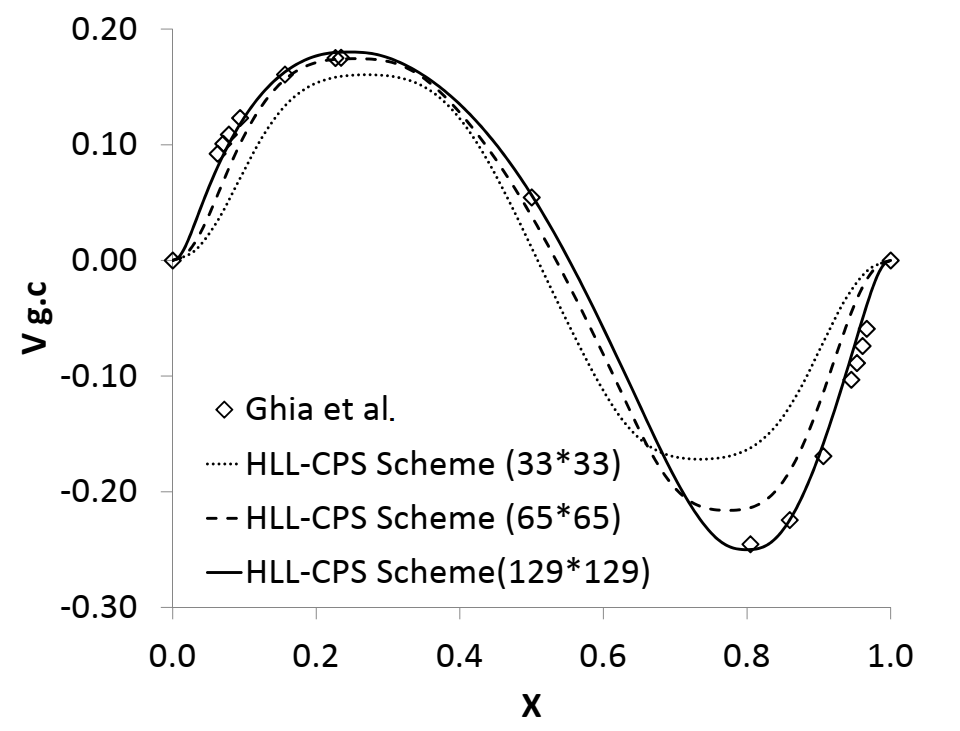}
	\caption{Velocity profile comparison of the HLL-CPS scheme for Re=100}
	\label{hllcps-re100}
\end{figure}

\begin{figure}[H]
	\includegraphics[width=200pt]{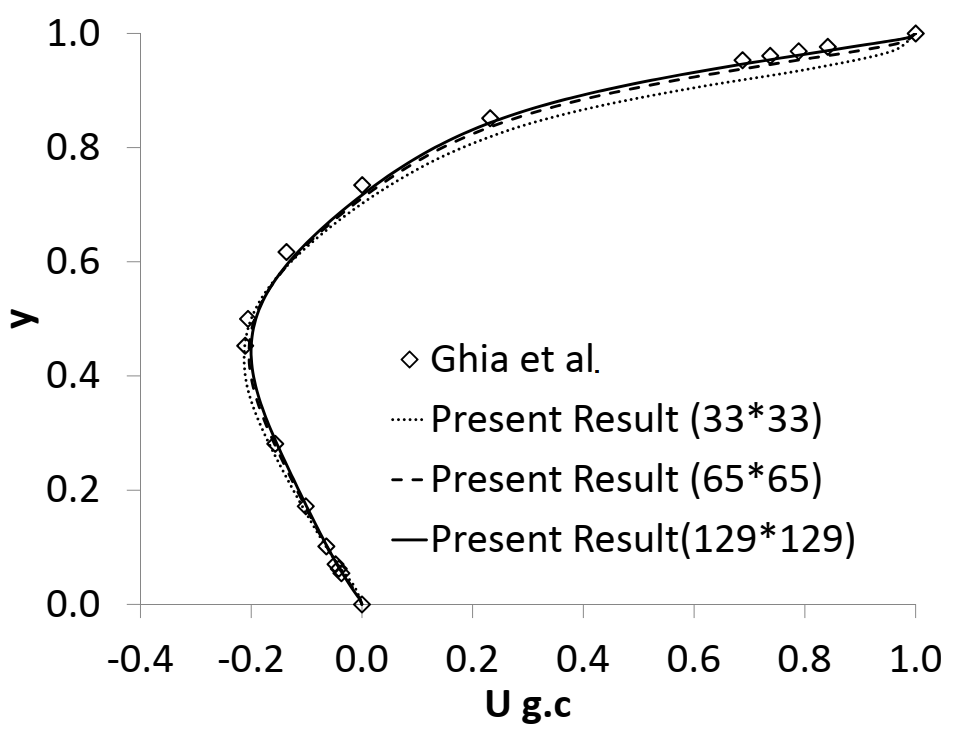}
	\includegraphics[width=200pt]{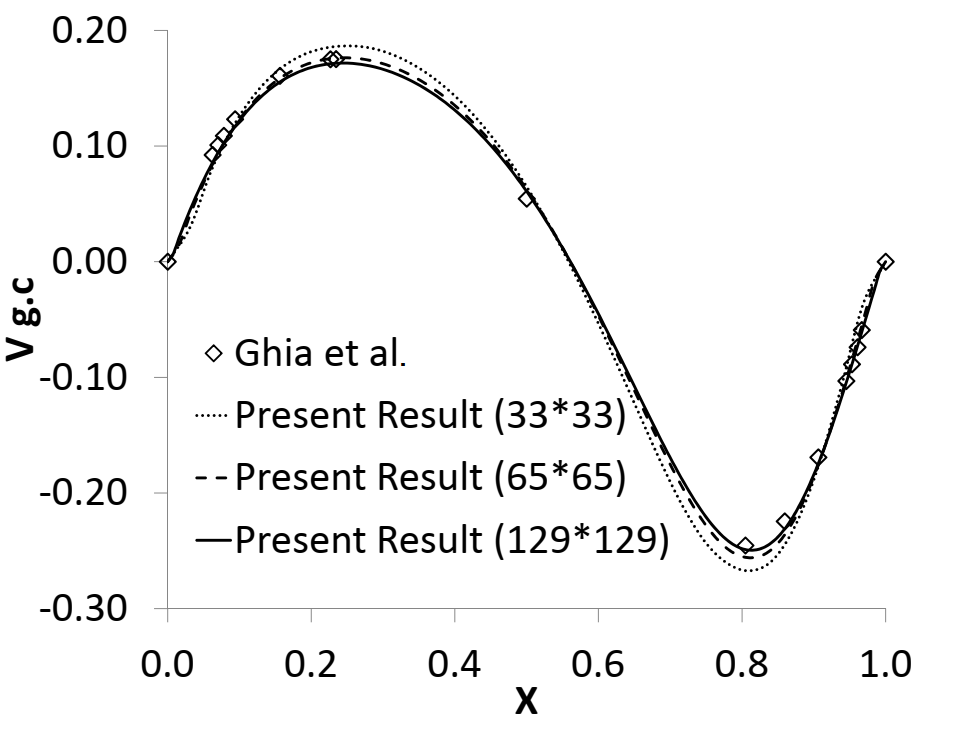}
	\caption{Velocity profile comparison of the HLL-CPS-RTNP scheme for Re=100}
		\label{hllcpsrtnp-re100}
\end{figure}

\begin{figure}
	\includegraphics[width=200pt]{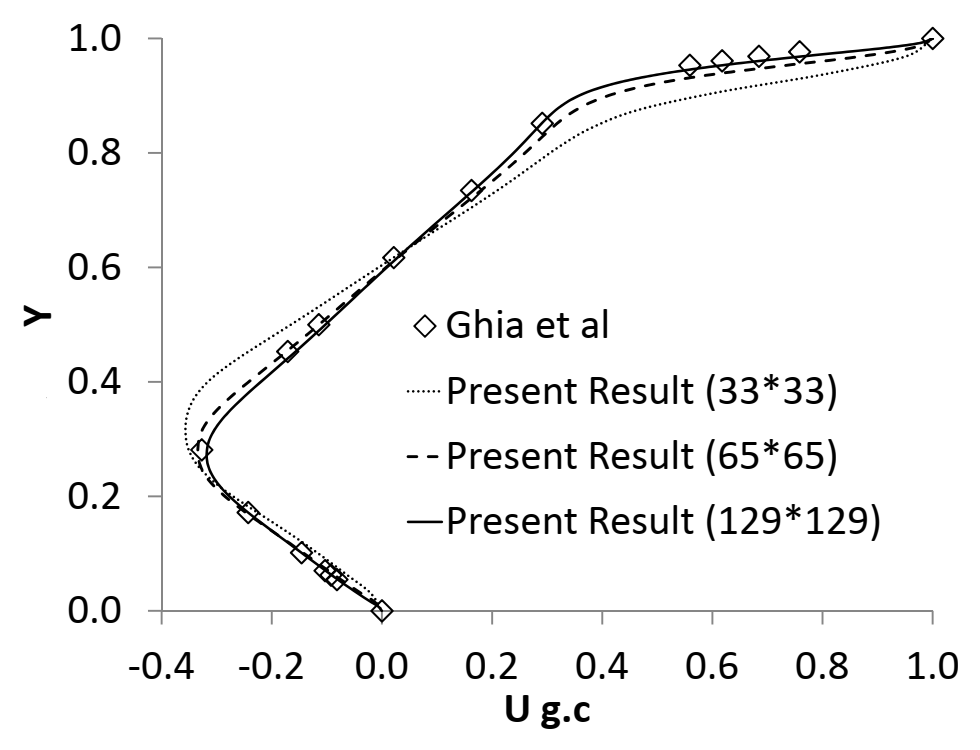}
	\includegraphics[width=200pt]{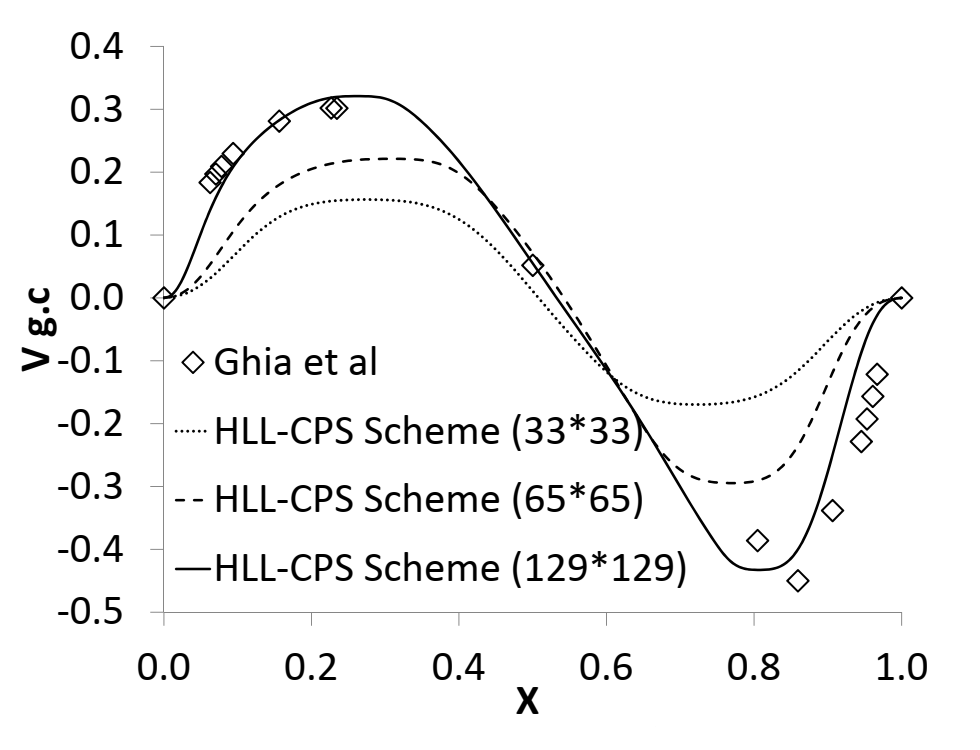}
	\caption{Velocity profile comparison of the HLL-CPS scheme for Re=400}
		\label{hllcps-re400}
\end{figure}

\begin{figure}[H]
	\includegraphics[width=200pt]{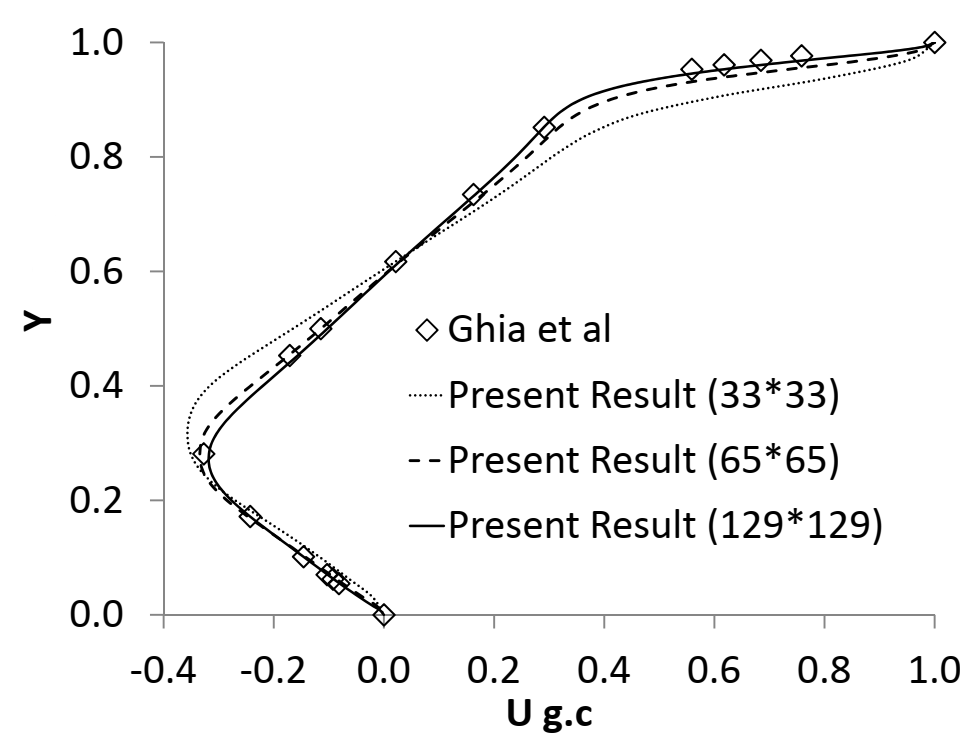}
	\includegraphics[width=200pt]{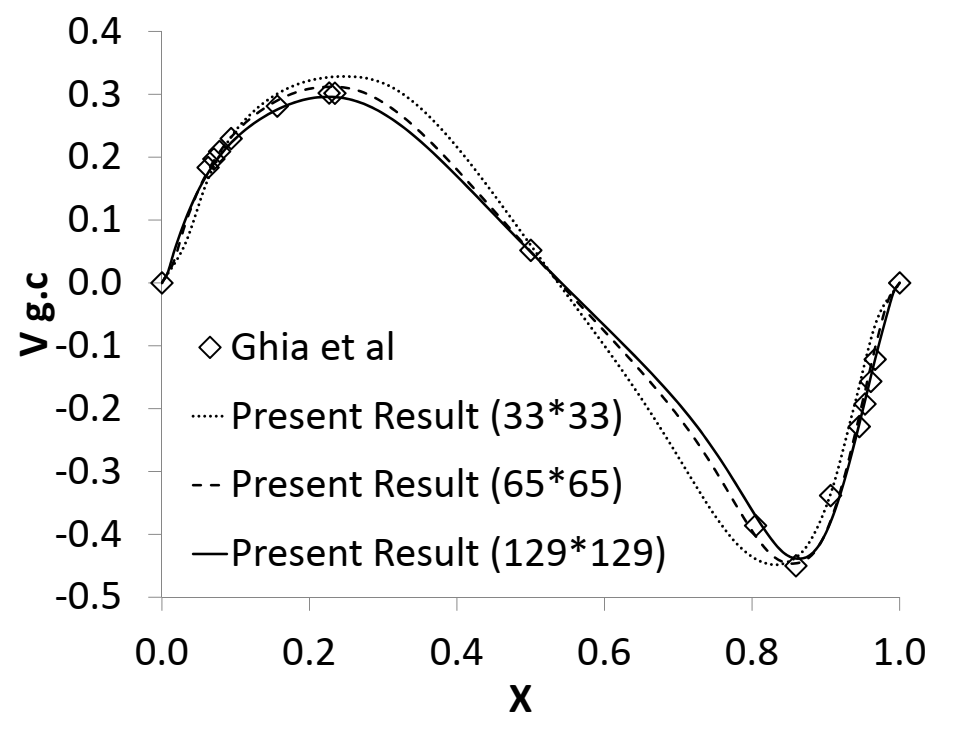}
	\caption{Velocity profile comparison of the HLL-CPS-RTNP scheme for Re=400}
		\label{hllcpsrtnp-re400}
\end{figure}

\begin{figure}[H]
	\includegraphics[width=200pt]{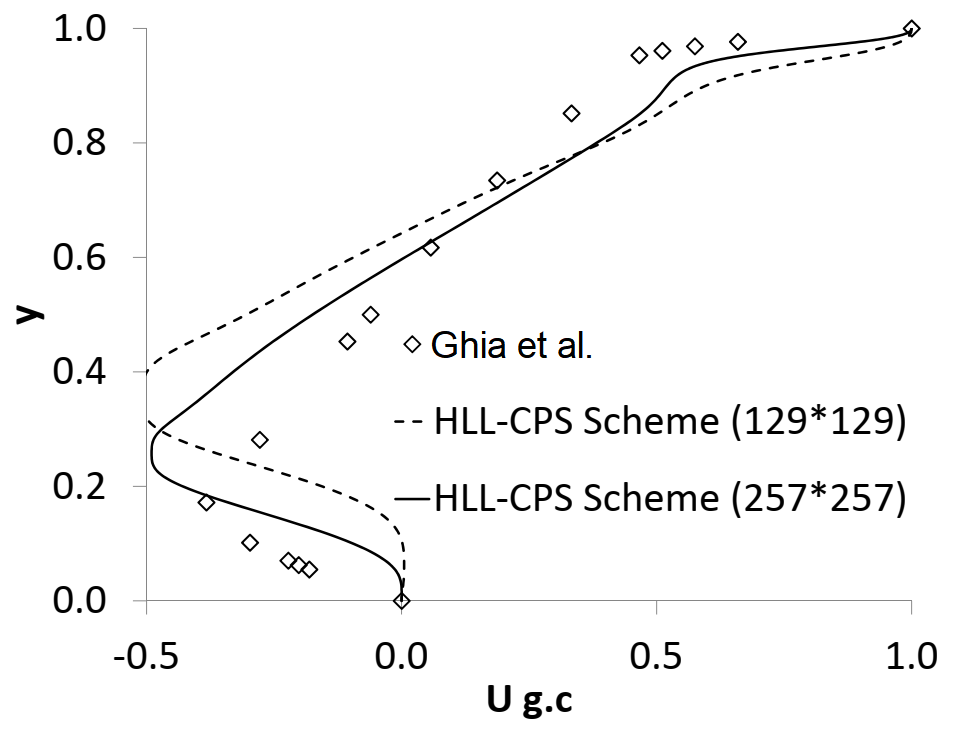}
	\includegraphics[width=200pt]{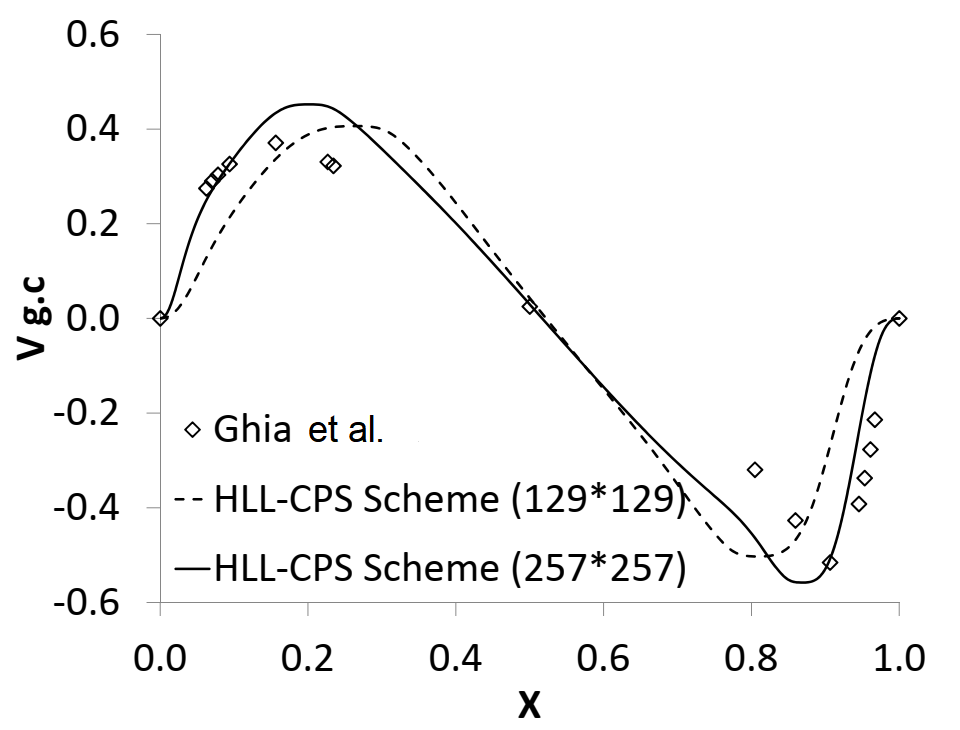}
	\caption{Velocity profile comparison of the HLL-CPS scheme for Re=1000}
		\label{hllcps-re1000}
\end{figure}

\begin{figure}[H]
	\includegraphics[width=200pt]{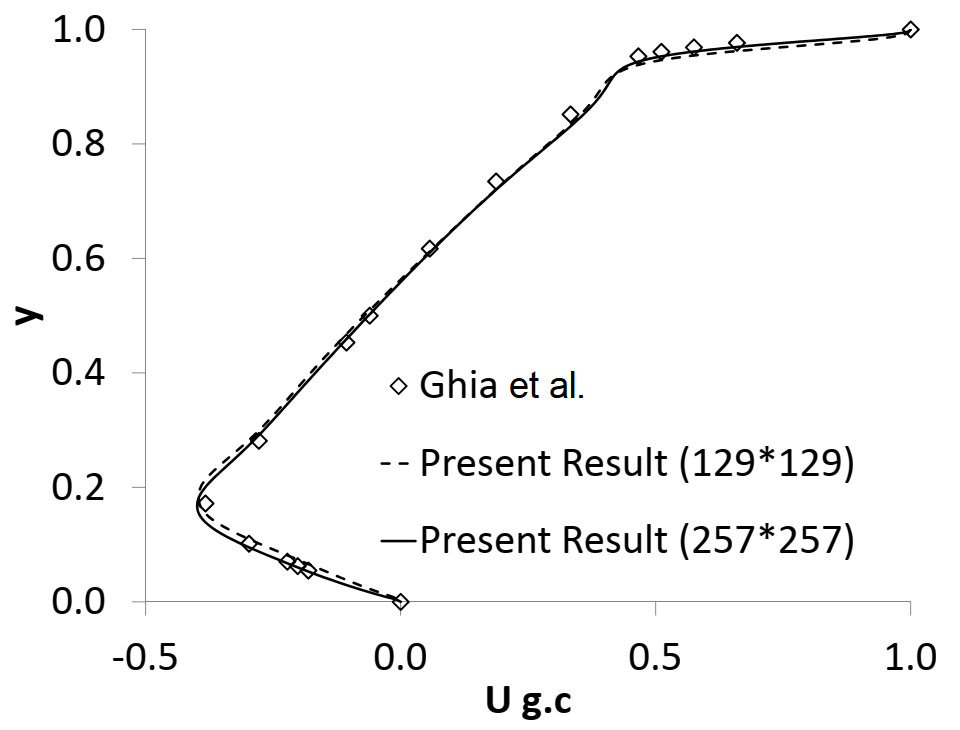}
	\includegraphics[width=200pt]{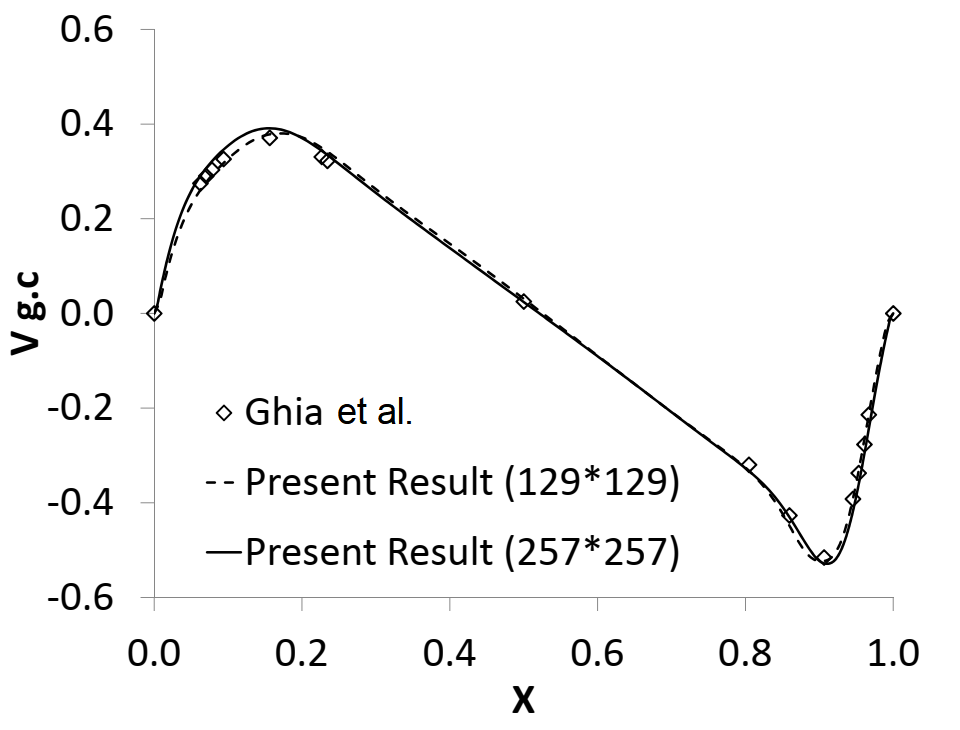}
	\caption{Velocity profile comparison of the HLL-CPS-RTNP scheme for Re=1000}
		\label{hllcpsrtnp-re1000}
\end{figure}

\begin{figure}[H]
	\includegraphics[width=200pt]{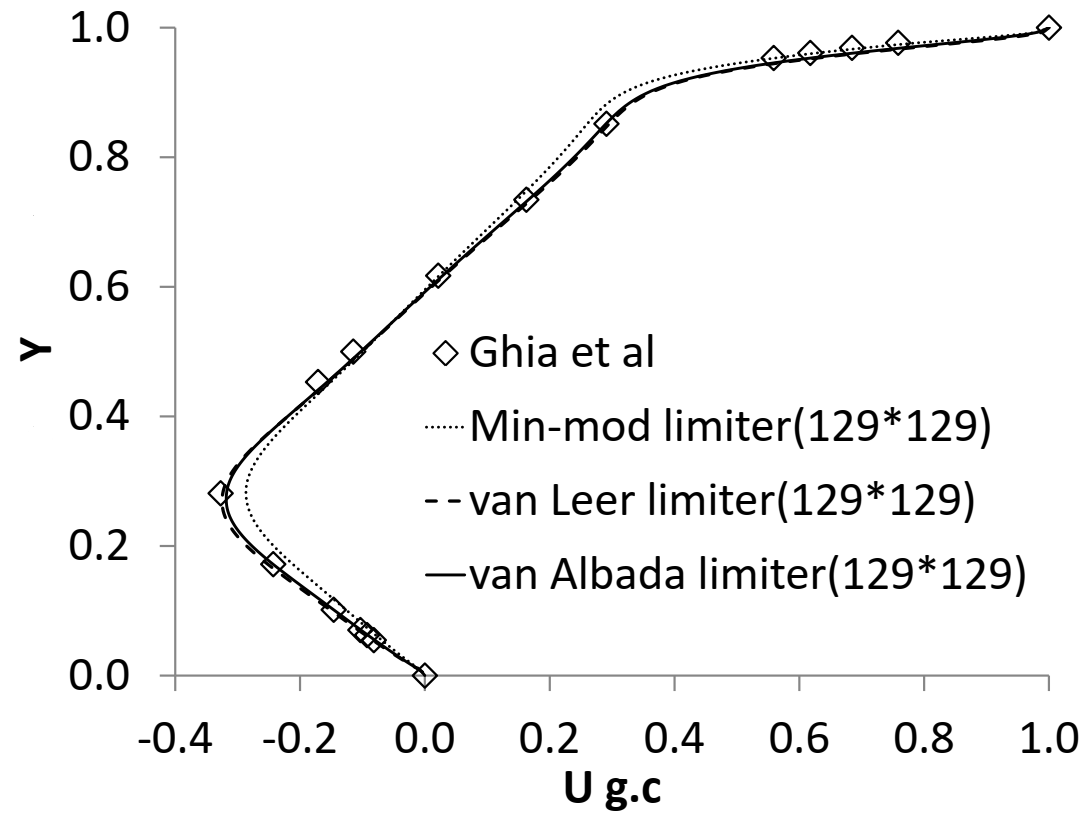}
	\includegraphics[width=200pt]{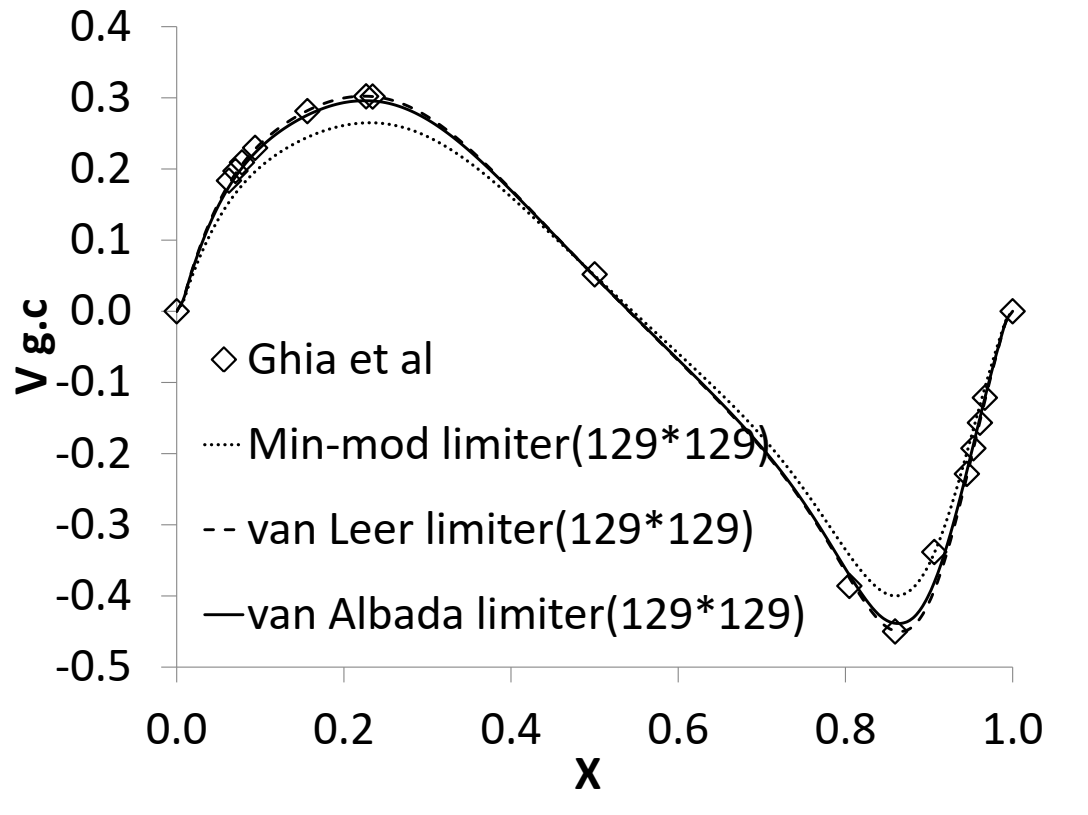}
	\caption{Velocity profile comparison of the HLL-CPS-RTNP scheme with different limiters for Re=400}
		\label{hllcpsrtnp-re400-limiter}
\end{figure}

The comparison of $u$ and $v$ velocity components along the geometric midsections with the results of Ghia et al. \cite{ghia} for a Reynolds number of 100 with the HLL-CPS scheme and the proposed scheme are shown in Fig. \ref{hllcps-re100} and \ref{hllcpsrtnp-re100} respectively. It can be seen from the figure that the velocity profiles of the HLL-CPS-RTNP scheme agree well with the results of Ghia et al. \cite{ghia} even on a coarse grid of 33$\times{}$33. For the HLL-CPS scheme, a grid size of 129$\times{}$129 is required for obtaining results of similar accuracy. The comparison of $u$ and $v$ velocity components along the geometric midsections with the results of Ghia et al. \cite{ghia} for Reynolds number of 400 are shown in Fig. \ref{hllcps-re400} and \ref{hllcpsrtnp-re400} respectively. It can be seen from the figure that a grid size of 65$\times$65 is required for obtaining accurate results with the HLL-CPS-RTNP scheme. For the HLL-CPS scheme, results are not at all close to the results of Ghia et al. even with a grid size of 129$\times{}$129. The comparison of $u$ and $v$ velocity components along the geometric midsections with the results of Ghia et al. \cite{ghia} for Reynolds number of 1000 are shown in Fig. \ref{hllcps-re1000} and \ref{hllcpsrtnp-re1000} respectively. It can be seen from the figure that a grid size of 129$\times{}$129 is required for obtaining accurate results with the proposed scheme whereas results of the HLL-CPS scheme are not at all close to the results of Ghia et al. The comparison of $u$ and $v$ velocity components along the geometric midsections with the results of Ghia et al. \cite{ghia} for Reynolds number of 400 using the HLL-CPS-RTNP scheme with different limiters is shown in Fig. \ref{hllcpsrtnp-re400-limiter}. The results with the van Albada and van Leer limiters are almost identical and match very well with the analytical results. The results with the min-mod limiter deviate marginally from the analytical results which may be due to the dissipative nature of the min-mod limiter.
\section{Conclusion}
A modified low diffusion HLL-CPS scheme, called HLL-CPS-RTNP, is proposed here for all Mach number flows. The stability of the original HLL-CPS scheme is improved  by introducing a diffusion of the contact wave in the vicinity of the shock. The stability of the proposed  scheme is demonstrated analytically by linear perturbation and matrix stability analyses. The proposed scheme is shown to be free from  numerical shock instabilities at high Mach numbers. Velocity reconstruction based on the face normal Mach number and a pressure function is proposed in the HLL-CPS scheme for improving  its ability further to resolve shear layers and low Mach flow features. With the face normal Mach number-based velocity reconstruction, the proposed HLL-CPS scheme resolves the shear wave accurately. For the cylinder case, the proposed scheme is found to be free from the inaccuracy problem of the original scheme and produces results resembling potential flow even at a Mach number of as low as 0.001. For lid-driven cavity cases, the proposed scheme is found to be accurate for a wide range of Reynolds numbers at very low Mach numbers. It is thus demonstrated that the proposed HLL-CPS-RTNP scheme is robust against numerical shock instabilities and it is capable of resolving shear layers and low Mach flow features very accurately.

\end{document}